%% file: paper.tex
\newif\iflongpaper
\longpapertrue
\pdfoutput=1

\documentclass{sig-alternate-10pt}

\input{setup}

\begin{document}

\author{
Stephanos Matsumoto\textsuperscript{\ddag\S} \;\;\;
Raphael M. Reischuk\textsuperscript{\S} \\[1mm]
Pawel Szalachowski\textsuperscript{\S} \;\;\;
Tiffany Hyun-Jin Kim\textsuperscript{\ddag} \;\;\;
Adrian Perrig\textsuperscript{\S}\\[4mm]
\textsuperscript{\ddag}CyLab/Carnegie Mellon University \qquad
\textsuperscript{\S}ETH Zurich
}

\title{Designing a Global Authentication Infrastructure (extended version)}

\maketitle

\begin{abstract}
  \input{abstract}

\end{abstract}

\section{Introduction}
\label{sec:introduction}\label{sec:intro}
\input{intro}

\section{Background}
\label{sec:background}
\input{background}

\section{Problem Definition}
\label{sec:problem}
\input{problem}

\section{\name Overview}
\label{sec:infrastructure}

\input{infrastructure}

\section{Isolation Domains}
\label{sec:ead}

\input{ead}

\section{Trust Roots}
\label{sec:trc}
\input{trc}

\section{Cross-Signing}
\label{sec:paths}
\input{paths}

\section{Separated Authentication}
\label{sec:separation}
\input{separation}

\section{Authentication Example}
\label{sec:validation}
\input{validation}

\section{Implementation and Evaluation}
\label{sec:evaluation}
\input{evaluation}

\section{Deploying \name}
\label{sec:deployment}
\input{deployment}

\section{Discussion}
\label{sec:discussion}
\input{discussion}

\section{Related Work}
\label{sec:related}
\input{related}

\section{Conclusions}
\label{sec:conclusion}\label{sec:concl}

\input{conclusion}

\section{Acknowledgments}
\input{acknowledgments}

\bibliographystyle{abbrv}
\footnotesize\bibliography{bib-truncated}

\appendix

\input{tec-dns}

\end{document}

%% file: setup.tex
\usepackage{amsmath}
\usepackage{xspace}
\usepackage{color,colortbl}
\usepackage[hyphens]{url}
\usepackage{mathptmx}
\usepackage{fancyvrb}
\usepackage{threeparttable}
\usepackage{rotating}
\usepackage{lastpage}
\usepackage{pifont}
\usepackage{boxedminipage}
\usepackage{paralist}
\usepackage{hyperref}
\usepackage{tabularx}
\usepackage[noadjust]{cite}

\newcommand{\dnssec}{DNSSEC\xspace}

\newcommand{\ead}{ISD\xspace}

\newcommand{\eads}{ISDs\xspace}

\newcommand{\shortcomings}{\textit{Shortcomings.}\xspace}

\newcommand\domain[1]{\texttt{#1}}

\newcommand\EID{EID\xspace}
\newcommand\EIDs{EIDs\xspace}
\newcommand\fullname{a Scalable Authentication Infrastructure for Next-generation Trust\xspace}
\newcommand\name{SAINT\xspace}
\newcommand{\mitm}{MitM\xspace}

\DeclareSymbolFont{symbols}{OMS}{cmsy}{m}{n}
\newcommand\cI{\ensuremath{\mathcal{I}}\xspace}
\newcommand\cJ{\ensuremath{\mathcal{J}}\xspace}
\newcommand\cK{\ensuremath{\mathcal{K}}\xspace}
\newcommand\cL{\ensuremath{\mathcal{L}}\xspace}
\newcommand\cM{\ensuremath{\mathcal{M}}\xspace}
\newcommand\cZ{\ensuremath{\mathcal{Z}}\xspace}

\newcommand{\myparagraph}[1]{{\medskip\noindent\textbf{#1.}}}

\newcommand{\note}[1]{}

\newif\ifnoindentaftersectionheadline
\noindentaftersectionheadlinetrue

%% file: abstract.tex
We address the problem of scaling authentication for naming, routing, and
end-entity certification to a global environment in which authentication
policies and users' sets of trust roots vary widely. The current mechanisms for
authenticating names (DNSSEC), routes (BGPSEC), and end-entity certificates
(TLS) do not support a coexistence of authentication policies, affect the entire
Internet when compromised, cannot update trust root information efficiently, and
do not provide users with the ability to make flexible trust decisions.  We
propose a Scalable Authentication Infrastructure for Next-generation Trust
(SAINT), which partitions the Internet into groups with common, local trust
roots, and isolates the effects of a compromised trust root. SAINT requires
groups with direct routing connections to cross-sign each other for
authentication purposes, allowing diverse authentication policies while keeping
all entities globally verifiable. SAINT makes trust root management a central
part of the network architecture, enabling trust root updates within seconds and
allowing users to make flexible trust decisions. SAINT operates without a
significant performance penalty and can be deployed alongside existing
infrastructures.

%% file: intro.tex
\ifnoindentaftersectionheadline\noindent\fi Alice lives in the Republic of
Mythuania and frequently travels around the world for business purposes. In her
business dealings, she frequently communicates with her clients and with her
banks, which are located all over the world. For the security of these
communications to client and bank websites, Alice primarily relies on three
mechanisms: DNSSEC~\cite{iana-ksk-url,dnssec-url} to authenticate
name-to-address mappings in DNS records, RPKI and
BGPSEC~\cite{rpki, bgpsec} to authenticate routes used to reach the sites,
and TLS~\cite{rfc5246} to authenticate the public keys used to establish a
confidential connection to the sites. These mechanisms rely on \emph{trust
roots}, which are assumed to be trustworthy, and follow one of two models:
\emph{monopoly} (the entire system has a single trust root) and \emph{oligarchy}
(users configure many trust roots of equal authority).

Unfortunately, these models both suffer from shortcomings that detrimentally
affect Alice's communication security. The monopoly model unrealistically
expects users to trust a single global root, such as ICANN for
DNSSEC. Alice must assume that ICANN correctly certifies DNS
records if she wants to have confidence in DNSSEC. As a Mythuanian citizen,
Alice may want to select a different set of trust roots from a citizen of
Oceania\footnote{This is the fictional state in George Orwell's novel
  \emph{1984}.}, but under the monopoly model neither citizen has a choice of
trust roots. Moreover, the global root is a single point of failure in the monopoly
model, which is particularly worrying in light of recent state-level
attacks~\iflongpaper
\cite{borger2013gchq, cremer2013denmark, rfc7258, gellman2013us,
greenwald2013how, weston2013snowden}.
\fi

\iflongpaper
  \begin{figure}[pt]
    \center
    \includegraphics[scale=0.6, trim=0 10 10 0]{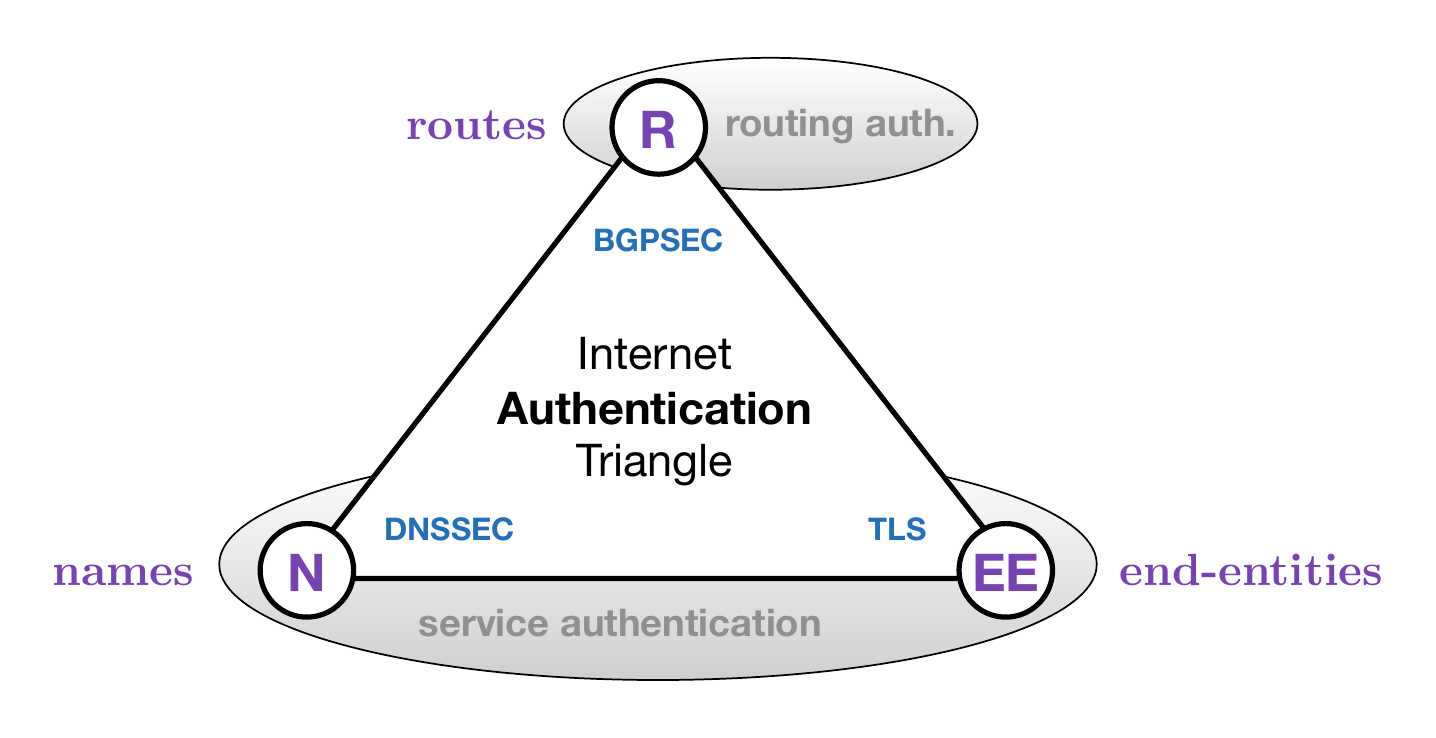}
    \caption{Authentication triangle for routes, names, and end-entities.}
    \label{fig:auth-triangle}
  \end{figure}
\fi

The oligarchy model, on the other hand, gives all trust roots equal, global
authority, allowing every trust root to certify information all over the world
(consider the example of root certificate authorities (CAs) in TLS).  However, this unfettered
global authority lacks an \emph{isolation} property: any compromised trust root
can affect authentication for any entity in the world, leading to
\emph{weakest-link security}. Under this model, Alice has difficulties
evaluating the trustworthiness of the many trust roots, preventing her from
effectively defending herself against compromised trust roots by ceasing to
trust them. Together with the inability to choose trust roots, we say that Alice
lacks \emph{trust agility}~\cite{marlinspike2011ssl}, the ability to select and
easily modify trust roots. Even if she can select trust roots in TLS, Alice
may cut off access to some sites by deselecting certain trust roots,
illustrating a need for \emph{global verifiability} with trust agility.

Because Alice travels frequently, she faces additional problems with today's
authentication mechanisms. Due to the fact that some trust roots in systems such
as TLS may only certify information in some parts of the world, Alice may have
to trust these roots when in the respective parts of the world, requiring her to
constantly change her set of trust roots. In contrast, she should have
\emph{trust mobility}, the ability to use the same set of trust roots wherever
she is in the world. To evaluate her confidence in a chain of signatures
authenticating a piece of information, she should have \emph{transparent
authentication}, knowing which trust roots are involved in certifying the
information.  Finally, in order to quickly obtain updated trust root information
in the face of a compromise or change, Alice should be able to learn of any
changes to her trust roots and obtain up-to-date information quickly and easily,
a property we call \emph{update efficiency}.

To address the above problems, we propose \fullname (\name), a series
of architectural design changes to current authentication mechanisms
(specifically DNSSEC, BGPSEC, and TLS).
Using Alice's example, we demonstrate how authentication in \name
provides stronger guarantees and additional properties over today's
Internet. In designing \name, we make the following contributions:
\begin{compactitem}

  \item We propose \textbf{isolation domains (ISDs)}, which partition the Internet into groups
    sharing common sets of trust roots, and scope trust root authority to these
    ISDs. The separation of ISDs allows Alice to select an ISD as
    a set of trust roots, enabling \emph{trust agility}, and the scoped authority of
    each ISD's trust roots enables \emph{isolated authentication}.

  \item We design \textbf{trust root configuration (TRC) files}, network messages containing trust root
    information. The distribution channel of these files is the same as that of
    routing messages, providing \emph{update efficiency}. Additionally, these files
    allow Alice to quickly obtain and use new trust root information, enabling
    \emph{trust agility}.

  \item We require \textbf{cross-signing} of trust roots between ISDs
    sharing a routing link using special certificates. This requirement enables
    \emph{global verifiability}, so that Alice can verify any entity to which
    she has a routing path. The certificates used for cross-signing allow Alice
    to see the ISDs through which she is authenticating her destination,
    enabling \emph{transparent authentication}.

  \item We separate authentication for \textbf{routing} and \textbf{service}
    entities, preventing circular dependencies in authenticating routes.  This separation also enables \emph{trust agility} and
    \emph{mobility}, allowing Alice's trust decisions for service entities to
    apply anywhere in the world.

\end{compactitem}

%% file: background.tex
\ifnoindentaftersectionheadline\noindent\fi
\iflongpaper
We provide a brief overview of existing authentication
infrastructures relevant to this work and their shortcomings.

\fi
\noindent\textbf{DNSSEC} \cite{iana-ksk-url,dnssec-url} was created to authenticate DNS
responses and thus to prevent cache poisoning and other attacks against DNS
security. ICANN operates the DNSSEC root signing key, which authenticates the
public keys associated with \domain{.com}, \domain{.org}, etc. In turn, these
keys authenticate the next level of the DNS hierarchy. %
Clients can authenticate the DNS responses by starting from the root key and
validating step-by-step the entire signature chain.

\shortcomings The single root of trust in DNSSEC implies that the entire world
is required to trust the root key, even though the world cannot agree on a
single trusted entity. Furthermore, despite the measures taken to protect the
root key from compromise, the key is still a single point of failure for
DNSSEC.

\medskip\textbf{BGPSEC}~\cite{LepTur2013} and
\textbf{RPKI}~\cite{rpki-url} constitute a standard to protect
BGP update messages against unwarranted modifications. BGPSEC
relies on RPKI for prefix and router certificates. RPKI
enables the authentication of AS numbers and IP address spaces.
In RPKI, each Regional Internet Registry (RIR) serves as a trust anchor and signs
certificates corresponding to resources, such as autonomous system (AS) numbers and
IP addresses, issued by that regional registry. For example, ARIN
signs a delegation for an address space provided to AT\&T, which in
turn signs a delegation to a customer of its subspace. The same
process occurs with AS numbers. Verifiers use the trust anchor
managed by each RIR to verify the delegation chain of
certificates for AS numbers and address spaces.
Before the owner of an address block advertises a prefix in BGPSEC,
it can use the address block certificate to sign a Route Origin
Authorization (ROA) to an initial AS. Each AS on the path
adds a signature of its own and the following AS number, called
a route attestation in S-BGP~\cite{KeLySe2000}. The route
attestations, together with AS number and address block certificates,
enable validation of the path in BGPSEC.

\shortcomings RPKI's validation process in BGPSEC suffers
from circular dependencies. To transfer routing information,
BGPSEC peers use the UPDATE message which contains signatures.
The certificate chains for the signature validation are stored at
each RIR's RPKI server. Hence, validation of the UPDATE message
requires each BGPSEC router to fetch the certificates directly
from the RIRs or from its local server, resulting in slow update
propagation.

\medskip\noindent\textbf{SSL/TLS} \cite{rfc5246} were created to secure web connections
between browsers and web servers. %
A web site is authenticated through an X.509 certificate that the
web site obtains from a Certification Authority (CA). Each browser
stores the public root key of each trusted CA, and uses one of these root keys
to validate a server's certificate. %

\shortcomings Numerous security issues exist with the key
infrastructure since current browsers trust around 650 root CA
keys~\cite{SSL-observatory}. %
Additionally, CAs have global jurisdiction and consequently any compromised CA
can issue a fake certificate for any server in the entire Internet. Recent
attacks on CAs
have underscored the fact that even
the most widely-used CAs suffer from such vulnerabilities, leading to
Man-in-the-Middle (\mitm) attacks on high-profile sites~\cite{Reischuk15:CaaI}.

\medskip\noindent\textbf{SCION} \cite{ZhHsHaChPeAn2011} is an isolation architecture for
inter-domain routing in the Internet. The provided isolation allows so-called
\emph{trust domains} (TDs) to distinguish between connections originating from inside or
outside the domain, and can guarantee that the path of communication between two
entities in a domain remains completely in that domain.
TDs are formed from ASes that are naturally grouped along
jurisdictional boundaries and who can agree on common roots of trust for routing
information. These boundaries protect misbehavior in one TD from affecting
routing in another TD.

\shortcomings SCION does not provide authentication for name lookups or end
entities, and does not specify mechanisms for the update or revocation of keys
in the routing architecture. Additionally, SCION ties users to the TD in
which they are located for all authentication, forcing them to trust their own
TD core (which serve as routing authorities) for all authentication.

Additional related work is discussed in \autoref{sec:related}.

%% file: problem.tex
\ifnoindentaftersectionheadline\noindent\fi
Our goal is to design global infrastructures allowing a user (Alice) to
authenticating routes, names, and EE certificates (such as TLS
certificates) for a server (Bob) in the Internet. In a global environment such
as the Internet, Alice does not trust all trust roots in the environment, and
she and Bob may not even have any trust roots in common. The trust roots may
differ in functionality and scope: some may authenticate routes and others
names, and some may be global and some local. We want to minimize global
authority, allow trust agility and mobility while maintaining global
verifiability, and allow Alice to evaluate the trustworthiness of information
being authenticated.

\myparagraph{Desired properties} In order to effectively the above
authentication problems, a network architecture should have the following
properties:

\begin{compactitem}

  \item \textbf{Isolated authentication.} The compromise of a trust root should
    be limited in scope. In particular, if Alice and Bob share trust roots for
    some information, no other trust root should be able to affect
    authentication of that information.

  \item \textbf{Trust agility.} Alice should have a clear, understandable choice
    over her trust roots. This choice should be easily modifiable at any time,
    with changes taking effect immediately (within seconds).

  \item \textbf{Trust mobility.} Alice's trust decisions should remain the same
    no matter where she is in the network. In other words, moving to different
    locations in the network should not require Alice to change her trust roots.

  \item \textbf{Global verifiability.} As in the current Internet, Alice should
    be able to authenticate any entity in the Internet that can be reached and
    has authentication information such as a name or EE certificate, even if its
    trust decision differs from hers.

  \item \textbf{Transparent authentication.} Alice should know when trust roots
    other than her own are certifying information that she verifies. In
    particular, for a chain of signatures Alice should be able to determine which
    trust roots are responsible for each signature.

  \item \textbf{Update efficiency.} Changes to trust root information (e.g., new
    keys and revocations) should take effect quickly (within minutes). In
    particular, Alice should be able to detect and obtain new information
    without requiring software updates.

\end{compactitem}
\myparagraph{Adversary model} Our adversary is an individual or organization
whose goal is to convince Alice of false information for a route, name, or
EE certificate.  To achieve this goal, the adversary can actively suppress,
change, replay, or inject messages into client-server communication, and might
also gain access to the private keys of trust roots in one domain. However,
besides these capabilities, the adversary cannot break cryptographic primitives
such as public-key encryption and hash functions.

\iflongpaper
\myparagraph{Other assumptions} In order for Alice to successfully verify
authentication information, she must also be able to verify a set of trusted
public keys which can then be used to bootstrap trust in other keys used during
authentication. We therefore assume that users like Alice
can verify an initial set of public keys through an out-of-band mechanism.
\fi

%% file: infrastructure.tex
\iflongpaper
\ifnoindentaftersectionheadline\noindent\fi
In this section, we highlight important features of \name. We provide
intuitive explanations of how these features accomplish the desired properties
mentioned in \autoref{sec:problem} and how they fit into the overall
\name architecture.
\fi

\subsection{Isolation Domains (ISDs)}
\label{sec:infrastructure:isd}

\begin{figure}[tp]
  \begin{center}
    \includegraphics[trim=0 50 0 10,width=0.9\linewidth]{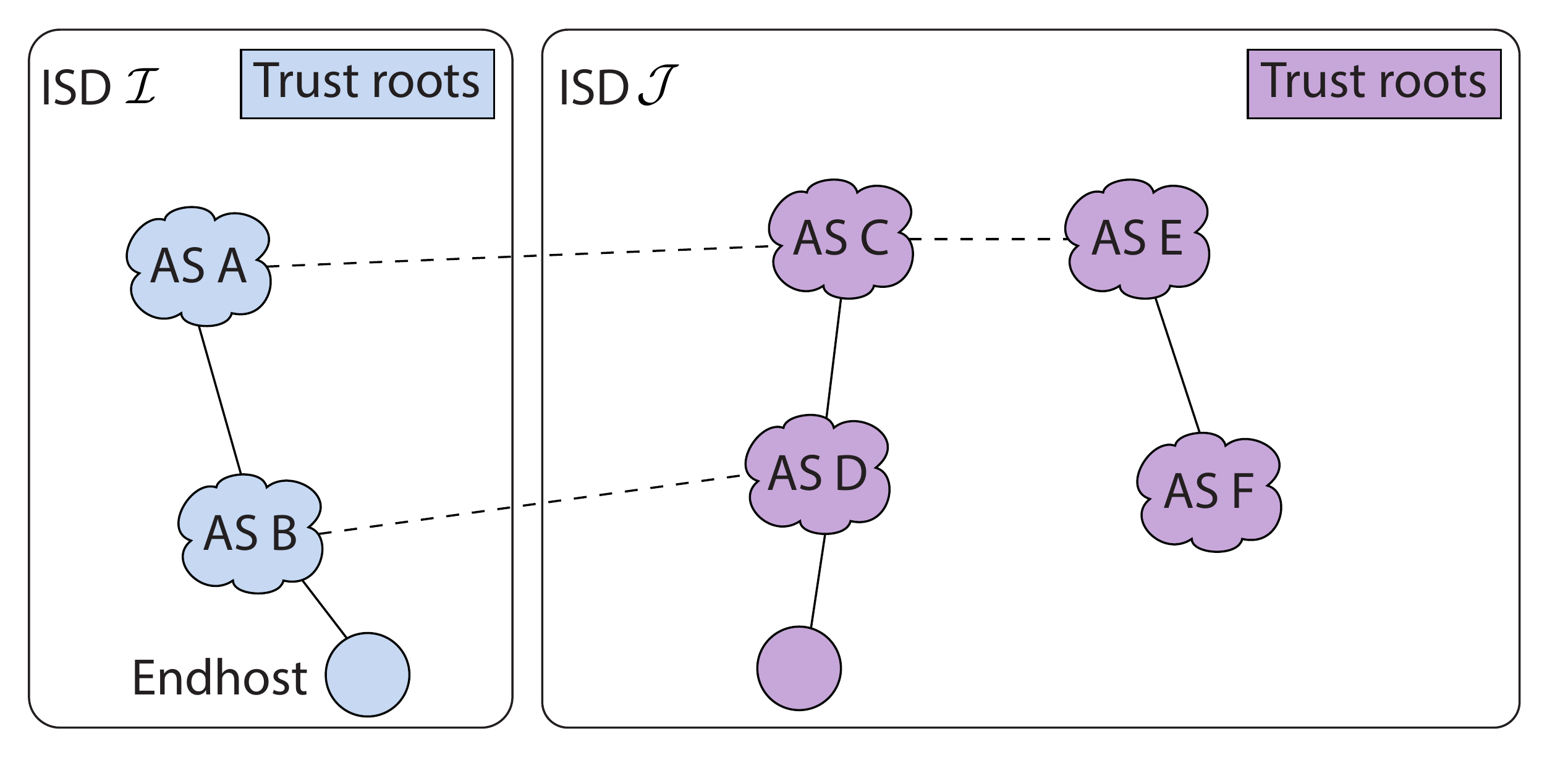}
  \end{center}
  \caption{Two isolation domains with individual trust roots.
  \iflongpaper The trust roots are illustrated in \autoref{fig:trustroots}. \fi
  The solid lines indicate provider-to-customer relations; the dashed lines
  indicate peering links.
  \vspace{-2mm}}
  \label{fig:ISDs}
\end{figure}

\noindent The Internet consists of a diverse assortment of groups, or domains,
each with its own set of trusted parties and individual policies regarding
routing, naming and EE certification. We make these differences a central
part of the \name architecture in order to achieve isolated authentication.
\name groups hosts,
routers, and networks into \emph{isolation domains} (ISDs), such as those shown in
\autoref{fig:ISDs}.
\name's ISDs are inspired by SCION's trust domains
\cite{ZhHsHaChPeAn2011} and leverage SCION's routing infrastructure.
However, \name's ISDs provide additional authentication mechanisms
for naming and EE certification.

An ISD is a collection of ASes with a common set of trust roots (see
\autoref{fig:trustroots}). These trust roots manage authentication within their
ISD, including management of routing, naming, and EE certification
policies, but are not authorities outside of the ISDs. The structure of
ISDs attempts to model existing trust relationships between humans by
grouping those with similar trust decisions together and by protecting users
from misconfigurations or breaches of trust outside these ``circles of trust''
where other trust decisions and policies hold. Thanks to ISDs, Alice can
select her roots of trust and when communicating within her own ISD, she
can stay protected from compromised trust roots outside of her ISD.
\autoref{sec:ead} describes the structure of ISDs in more detail.

In practice, ISDs can represent groups of various scales, such as companies,
conglomerates, or countries. ISD-level policies will vary greatly by the scale
of the ISD, since corporate policies often contain much more detail than
country-wide laws.  In this paper we use countries as examples of ISDs for
several reasons: (1) international boundaries approximately map to DNS naming
boundaries, which are also separated in \name, (2) national data privacy laws
provide a reasonable example of security policies in top-level ISDs, and (3) the
resulting set of domains represents an easily-understood choice among possible
sets of trust roots, since users can more easily understand what it means to
evaluate and trust a country (representing a set of trust roots) rather than
doing so with individual trust roots.

\subsection{Trust Root Configuration (TRC) Files}
\label{sec:infrastructure:trc}

\noindent Trust root management in \name is handled by TRC files, which
contain information about an ISD's trust roots, such as their public keys
(see \autoref{sec:trc:trc} for more details). TRC files are disseminated as
network messages along the same channels as routing messages and DNS
responses (see \autoref{sec:trc:distribution}), providing update efficiency.
Because routing messages are required to maintain connectivity, new TRC
files can quickly propagate throughout the network in case a trust root is
compromised. This mechanism allows Alice to quickly obtain up-to-date trust root
information.

In addition to the above distribution mechanisms, TRC files can also be
downloaded and chosen by the users as a new set of trust roots. Since a TRC
file contains all of the necessary trust root information, a user like Alice can
easily switch to a different set of trust roots by simply obtaining and
selecting a different TRC file. In essence, \name provides trust
agility by allowing users to easily modify their trust decisions at any time.

TRC files contain trust root information for a given ISD and thereby
enable transparent authentication. Namely, when a trust root signs the
information of another ISD's trust root (as explained below), it does so by
signing the TRC file of the other ISD. Thus a chain of signatures
clearly indicates domain boundaries by design. Alice can use this knowledge of
ISD boundaries to evaluate the trustworthiness of this signature chain and
determine whether or not to accept the authenticated information.

\subsection{Cross-Signing Trust Roots}

\noindent If we consider ISDs as countries, then we can clearly see that
not all ISDs' trust roots will cross-certify one another, and it is
unrealistic to expect countries to do so. Rather, we only require the trust
roots of two ISDs to cross-certify one another if they share routing links,
that is, if they are physically connected and route traffic through one another.
This requirement ensures that the existence of a routing path implies the
existence of a chain of signatures for a name or EE certificate, providing
global verifiability by allowing Alice to verify this information for any entity
she can reach (see \autoref{sec:paths} for more details).

This cross-signing requirement also helps to provide mobility: no matter
where users are located, they can authenticate service information (names and EE
certificates) starting from their own trust roots (named in their ``home''
TRC file) to the ISD of the entity whose information they are
verifying. Thus as long as Alice can reach her home ISD from the ISD
in which she is located, she can use her existing trust decision for
authentication anywhere in the world.

\subsection{Separation of Authentication Types}
\label{sec:infrastructure:separation}

\noindent
\name separates routing authentication from service authentication
(which certifies names and EE certificates). Because authentic routes are
required to fetch necessary information during name lookups and EE certificate
handshakes, we treat routing as a separate authentication mechanism. Moreover,
we note that authentication of routes cannot rely on fetching external
information, as this would itself require authentic routes and thus create a
circular dependency.

The separation of routing and service authentication also helps provide trust
mobility in \name. We observe that users' physical locations indeed
influence their routing authentication; in particular, a route from Alice to Bob
must be authenticated by trust roots of the ISDs in which Alice and Bob are
located.  However, this requirement does not hold for service authentication;
thus Alice can use the trust roots of an ISD of her choosing to completely
bypass the ISD in which she is located to authenticate names or EE
certificates, providing trust mobility and greater resilience against \mitm
attacks.

%% file: ead.tex
\iflongpaper
\ifnoindentaftersectionheadline\noindent\fi
We now discuss isolation domains in more detail. We begin by
describing the physical layout of \eads along with the structure of
the namespace and the address space in \name. We then present the
concept of trust anchor \eads, which provide starting points for the
authentication of routes, names, and EE certificates.
\fi

\subsection{ISD Structure}

\noindent
An ISDs is made up of multiple networks, or ASes, as shown in
\autoref{fig:ISDs}, with ISDs connecting to one another to enable
Internet-wide connectivity. Similarly to trust domains in SCION, \name
arranges ASes hierarchically within an ISD by customer-provider
relationships. The ASes in the top tier (those with no providers) are
referred to as the \emph{ISD core} and serve as the trust roots for routing
(see \autoref{fig:trustroots} and \autoref{sec:trc:roots} for more details).

The connectivity of ASes in \name is similar to the current BGP-based
relations: \eads are primarily connected by links between \ead core ASes
(similar to BGP transit links between tier-1 ISPs), but can also be connected by
links between lower-tier ASes (similar to BGP provider-customer links and
peering links).
Paths between \ead cores are determined by a simple routing protocol among the
core ASes of each \ead. Based on our analysis of the current Internet
(\autoref{sec:deployment}), we expect that there will be on the order of several
hundred core ASes in total. Therefore, in running such a protocol we do not
worry about significant overhead or scalability.

\subsection{DNS Namespace}
\label{sec:ead:namespace}

\begin{figure}[t]
  \centering
  \includegraphics[trim=0 20 0 0, width=0.9\linewidth]{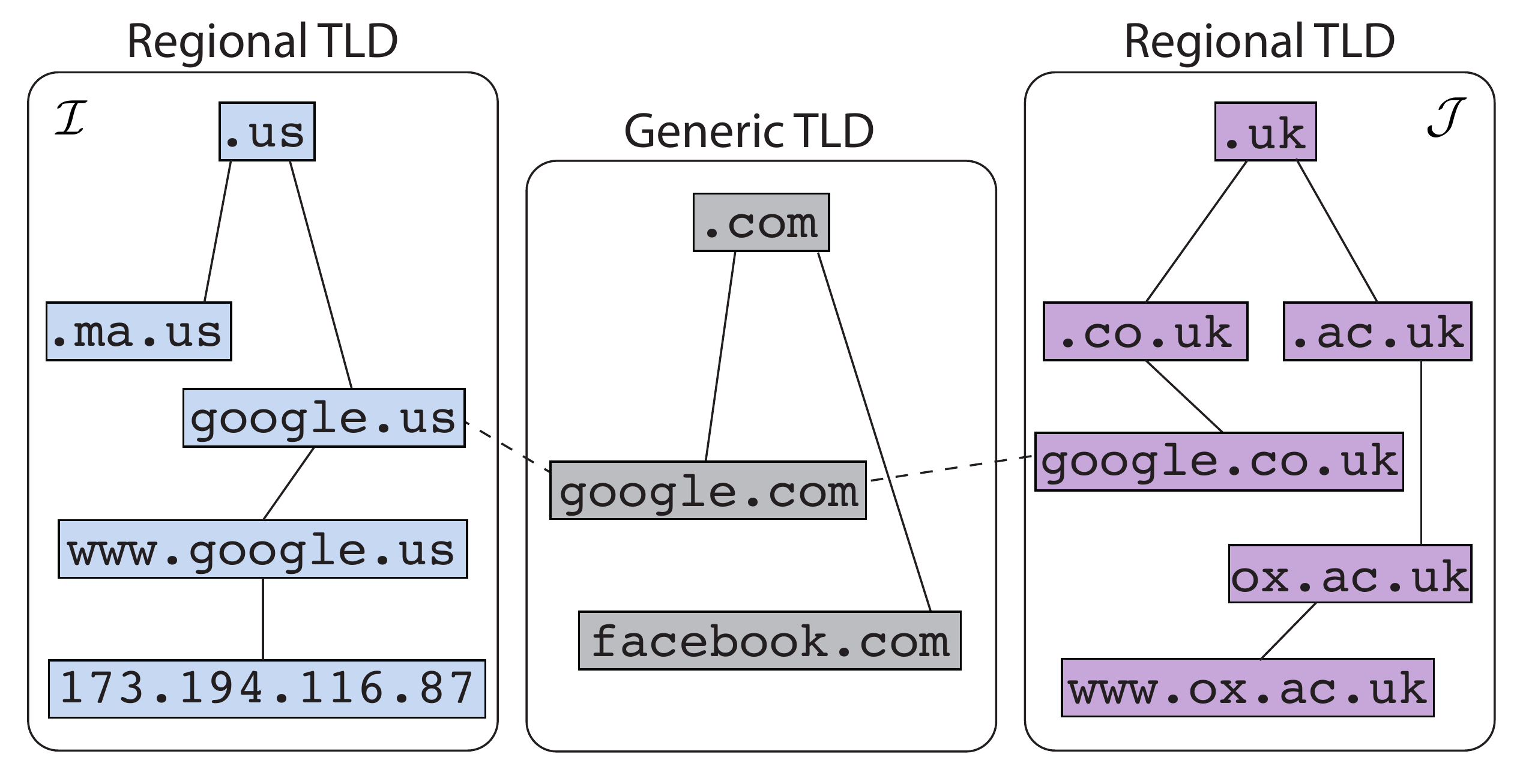}
  \caption{Namespace structure in \name. Solid lines indicate hierarchical
  relations, and dashed lines indicate redirections.}
  \label{fig:isd:namespace}
\end{figure}

\noindent
Each \ead has the autonomy to manage its own namespace. We structure
\name's global namespace as a collection of top-level domains (TLDs)
bound by an inter-domain web of trust~\cite{zimmerman1994pgp} (as
shown in \autoref{fig:isd:namespace}), rather than by a global root
zone as is done in DNSSEC. However, \name's name resolution process
is similar to that of DNSSEC.

Each \name DNS root server answers queries for hosts within its ISD.
An ISD's namespace supports one of two top-level domain types:
\begin{compactitem}

  \item \textbf{Regional TLDs} in \name correspond to a specific \ead. In the
    example of \autoref{fig:isd:namespace}, the TLD \domain{.us} represents the
    United States ISD, whereas \domain{.uk} represents the United Kingdom
    ISD. In order to provide transparency, the DNS server responsible for a
    regional TLD guarantees that any address record (similar to A records in
    DNS) maps to an address within the corresponding ISD.

  \item \textbf{Generic TLDs} such as \texttt{.com} and \texttt{.org}, by
    contrast, are \emph{not} bound to any specific ISD, and can thus name an
    entity located \emph{anywhere} in the world. However, a name in a generic
    TLD can only map a redirection to another name, thereby ensuring that only
    names under regional TLDs map to addresses (and only within the TLD's
    corresponding ISD). This guarantee provides domain transparency during
    DNS lookups.
    \iflongpaper
    Details about name resolution for generic TLDs are in
    Appendix~\ref{sec:tec:dns:gTLDs}.
    \fi

\end{compactitem}

We expect that today's ccTLDs such as \domain{.us} and \domain{.uk} will
continue to operate as regional TLDs under \name. Countries such as Tuvalu
(whose ccTLD is the popular \domain{.tv}) may choose to operate as a generic TLD
and continue to sell names that map all over the world, but must do so through
redirections to other names.

\subsection{Address Space}
\label{sec:ead:membership}

\noindent We define an address in \name as a 3-tuple of the form $(\cI, A, e$),
where $\cI$ represents an ISD identifier, $A$ represents an \emph{AS identifier} (ASID),
and $e$ represents an \emph{endpoint identifier} \EID. For example, if Alice wants to reach Bob,
who has the \EID 42ac6d in AS 567 and ISD $\cI$, she would contact the
address $(\cI, 567, 42\textrm{ac}6\textrm{d})$.
In contrast to the current Internet, AS numbers and \EIDs do not have
significance outside of their respective ISDs and ASes, and thus can
have any format. An \EID, for example, can be an IPv4, IPv6, MAC, or
self-certifying address.

\emph{Registry servers} in ISDs (described in \autoref{sec:trc:roots})
assign ASIDs to ASes, and similar servers in each AS issue
\EIDs to endhosts.  Due to the local significance of ASIDs and
\EIDs, $(\cI, A, e)$ and $(\cJ, A, e)$ are distinct addresses even though
both have the same ASID and \EID. This addressing scheme allows full
address to be globally unique while giving ISDs and ASes the autonomy
to manage addressing as well as names within their own realm of control.

\iflongpaper
The above addressing system also allows for interoperability with the current IP
addressing and AS numbering schemes while simultaneously enabling local
deployments of other proposed addressing schemes. For example, some ISDs may
choose to retain the current AS numbering and IP addressing schemes, while
others may opt to provide ASes with human-readable identifiers and endpoints
with IPv6 addresses.

Similarly to ASIDs, endpoint addresses (since only used locally in intra-AS
routing) do not need to be globally allocated like the current IP
address space. Since the routing authentication infrastructure only handles
inter-AS routes, \name does not bind the endhost address space to public keys as
RPKI does. Instead, forwarding from an edge router to an endpoint address is
resolved and handled entirely within an AS. \note{could mention IPv4 running
out of addresses here, but it doesn't fit well with the paragraph}
\fi

\subsection{Trust Anchor \eads}
\label{sec:ead:anchor}

\noindent Trust anchors in the current Internet, such as IANA for
RPKI and BGPSEC, ICANN for DNSSEC, and root CAs in
TLS, represent starting points for authenticating information. Similarly,
trust anchor ISDs are starting points for authenticating routes, names, and
EE certification in \name. Due to the separation of authentication by
type, Alice can anchor her trust for authenticating routing and service
information in separate ISDs.

As discussed in \autoref{sec:infrastructure:separation}, for routing purposes
the trust roots of the ISD's in which Alice is currently located must
certify all of her routes. However, Alice can select the trust roots of any
ISD to authenticate service information (names and EE certificates).
We call this ISD Alice's \emph{trust anchor ISD}. Alice choice of this
ISD can be easily changed at any time (as described in
\autoref{sec:trc:distribution}). An example of such trust bootstrapping is
provided in \autoref{sec:validation}.

%% file: trc.tex
\iflongpaper
\ifnoindentaftersectionheadline\noindent\fi
In this section, we cover what entities serve as trust roots and how trust roots
are configured for an ISD. We also discuss how we update trust root
information using network-level messages, and how this incorporation of trust
management into the network allows for fast updates of trust root information.
Finally, we describe our scheme of separating trust roots by ISD, and how
separated categories of authentication enable trust agility.
\fi

\subsection{Trusted Parties}
\label{sec:trc:roots}

\begin{figure}[t]
  \centering
  \includegraphics[trim=0 10 0 0, width=0.9\linewidth]{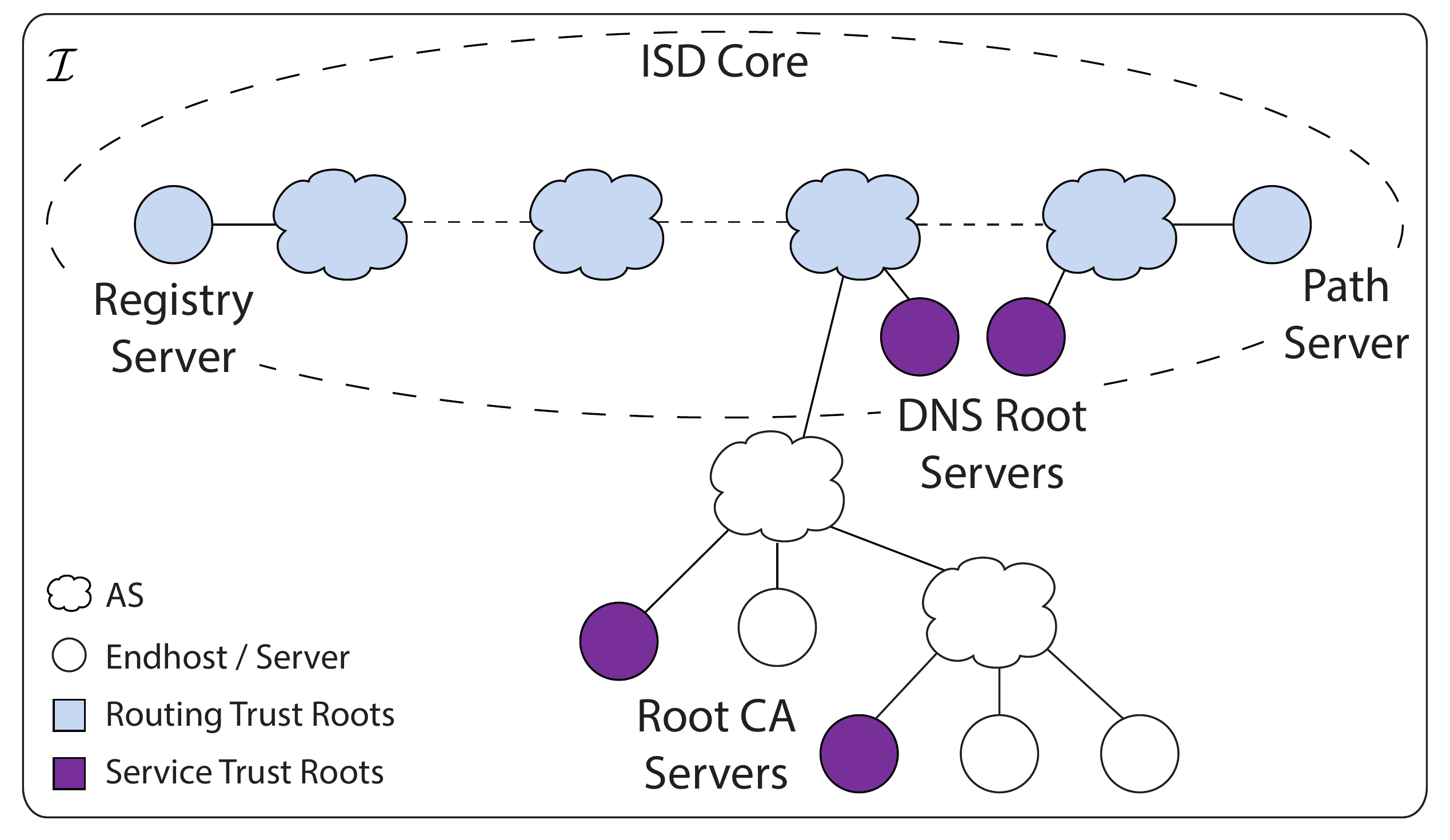}
  \caption{Logical and physical placements of trust roots\iflongpaper in an ISD\fi.}
  \label{fig:trustroots}
  \label{fig:ASregistration}
\end{figure}

\noindent Trust roots sign authentication information for routes, names, or
EE certificates, and set policies governing the ISD. These policies
may include information such as preferences for certain encryption or signature
algorithms or constraints on certificate validity. A trust root is an authority
for either routing or service authentication (see \autoref{fig:trustroots}).

Because a trust root is only an authority in its own ISD, a compromised
trust root cannot enable the impersonation of servers in other ISDs.
Scoping trust root authority to the ISD level protects Alice and Bob's
communication from many compromises in other ISDs and provides guarantees
to Alice about authentication in her trust root ISD.

We classify trust roots into routing and service trust roots. The
\textbf{routing trust roots} consist of the following parties:
\begin{compactitem}

  \item The \emph{core ISPs} are responsible for sending out route
    announcements, which are propagated from providers to customers and
    establish cryptographically signed paths from the recipient to the core.

  \item The \emph{path server} stores and provides a lookup service for mappings
    between an ASID and the AS's down-paths. These paths are registered by
    the ASes at the core. The path server is co-located with and operated by the
    core ISPs.

  \item The \emph{registry server} issues and stores \emph{AS certificates}
    binding an ASID to its public key (called its \emph{AS key}), which are
    used to verify the signed paths provided by the path server. Like the path
    server, the registry server is co-located with and operated by the core
    ISPs.

\end{compactitem}

The \textbf{service trust roots} consist of the trust roots for naming and for
EE certification. The \emph{DNS root} is the starting point for
verifying all names in the ISD's namespace. The DNS root also sets
ISD-wide naming policies such as reserved or forbidden domain names and
allowed signature algorithms to sign records. Because the failure of the
DNS root can block user connectivity in an ISD, the DNS root
should be highly robust and available, using mechanisms such as distributed
anycast schemes and placing servers in the ISD core where they can be
reached through highly available top-tier ASes.

The \emph{root CAs} are the starting points for verifying EE
certification information in an ISD. Root CAs in \name serve the same
purpose as they do in today's PKIs by signing TLS public-key certificates.
However, they are restricted to only signing EE certificates in ISDs in
which they are root CAs. They can also sign intermediate CA certificates as in
today's TLS PKI. If the ISD uses other public-key infrastructures such as CT or
AKI (see \autoref{sec:discussion}), then the trust roots for EE
certification also include public logs and auditors/validators.

\subsection{Trust Root Configuration (TRC) Files}
\label{sec:trc:trc}

\noindent
As mentioned in \autoref{sec:infrastructure:trc}, a TRC file
specifies the trust roots for an ISD, the public keys of those trust roots,
and the authentication policies of the ISD. It also specifies the locations
of the DNS root and TRC servers (described in
\autoref{sec:trc:distribution}) to allow users to reach these servers without
performing DNS lookups. TRCs are created and managed by an ISD's trust
roots and distributed through the routing mechanism. Specifically, a threshold
of trust roots is required to sign a new or updated TRC file, and the core
ISPs distribute the TRC file within the ISD through a
broadcast
mechanism that we describe below.

The quorum of trust roots required to update the TRC file is specified in
the TRC file itself, providing the trust roots with the autonomy to set
their own threshold for altering the TRC file. A higher threshold is more
secure to a compromise of multiple trust roots, but also reduces the efficiency
in updating TRC files. The TRC file also specifies a quorum of trust
roots that must sign a cross-signing certificate to authenticate another
ISD's trust roots.  Cross-signing certificates are described in more detail
in \autoref{sec:paths}.

\myparagraph{TRC format} A TRC file is encoded as an XML file with the
fields shown in \autoref{tab:TRC}. The version number and timestamp ensure that
users can verify information using recent policies and trust root
information. The public keys of the ISD's trust roots provide starting
points for verifying routes, names, and EE certificates.
The TRC file may also contain the public
keys of additional entities (e.g., public logs in CT~\cite{laurie2013certificate} or
AKI~\cite{kim2013accountable}).
In order to allow users to easily reach the DNS root of an ISD, the
TRC also contains one or more addresses for the ISD's DNS root.

\begin{figure}[bt]
\begin{center}
  \scriptsize
  \begin{tabularx}{\linewidth}{|c|X|}\hline
Field & Description \\\hline \hline
\texttt{isd} & ISD identifier \\\hline
\texttt{version} & Version of TRC file \\\hline
\texttt{time}& Timestamp \\\hline
\texttt{coreISPs}& List of core ISPs and their public keys \\\hline
\texttt{registryKey}& Root registry server's public key \\\hline
\texttt{pathKey}& Path server's public key \\\hline
\texttt{rootCAs} & List of root CAs and their public keys\\\hline
\texttt{rootDNSkey} & DNS root's public key  \\\hline
\texttt{rootDNSaddr} & DNS root's address \\\hline
\texttt{trcServer} & TRC server's address \\\hline
\texttt{quorum} & Number of trust roots that must sign new TRC \\\hline
\texttt{trcQuorum} & Number of trust roots that must sign an ISD cross-signing cert \\\hline
\texttt{policies} & Additional management policies for the ISD \\\hline %
\texttt{signatures} & Signatures by a quorum of trust roots \\\hline
\end{tabularx}
\end{center}
\vspace{-2mm}
\caption{Fields in a TRC file.}%
\label{tab:TRC}
\end{figure}

\myparagraph{Policies} A TRC file can also specify additional policies
related to ISD management. For example, these policies might specify a
minimum key length or required encryption algorithms for all EE certificates in
the ISD. Systems such as PoliCert~\cite{szalachowski2014policert} have
proposed similar policies on a per-domain basis; we leave a detailed design of
additional ISD-wide policies to future work.

\myparagraph{Updating the TRC file}
In the event that a TRC file needs to be updated, the trust roots confer
out of band to determine what changes need to be made to the TRC file. Once
they have decided to update a TRC file, the trust roots sign the new
TRC file. Each of these signatures is appended to the \texttt{signatures}
section of the new TRC file, and sent when a quorum of trust roots signs
the~TRC file.
\iflongpaper
The trust roots can also use group
signatures~\cite{chaum1991group} or threshold
signatures~\cite{shoup2000practical} to update the TRC file.
\fi

\subsection{TRC Distribution and Management}
\label{sec:trc:distribution}

\iflongpaper
\begin{figure}[t]
  \centering
  \includegraphics[trim=0 15 0 0,width=0.9\linewidth]{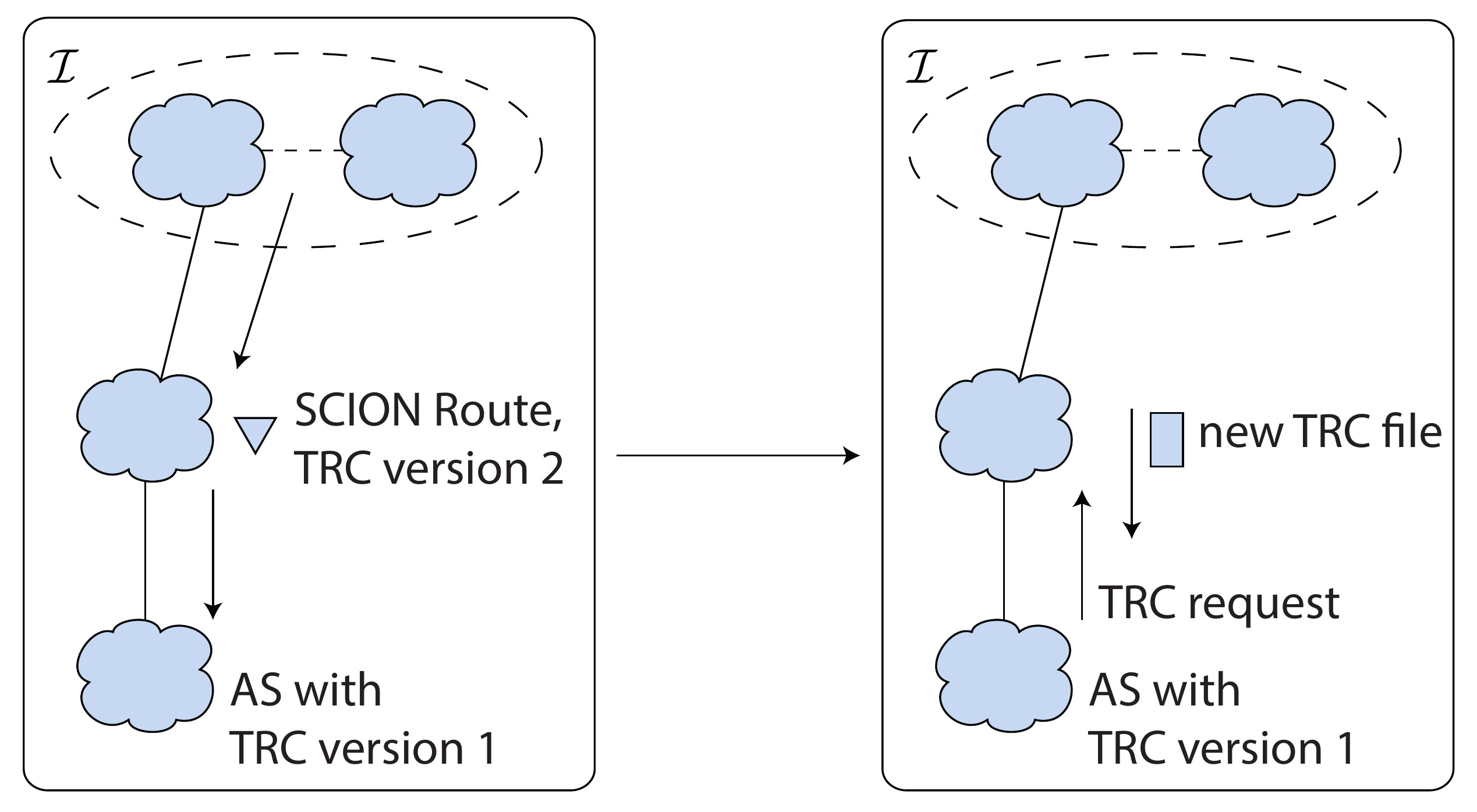}
  \caption{Distribution mechanism for updated TRCs. Arrows indicate sent
  network messages.}
  \label{fig:update}
\end{figure}
\fi

\noindent \textbf{Obtaining an initial TRC file.} We envision that Alice will
most commonly obtain a initial TRC file of her provider's ISD when forming a
service agreement.  If Alice wants to obtain a different ISD's TRC file, she can
contact the \emph{TRC server} of that ISD, a server that stores the TRC files
and cross-signing certificates (see \autoref{sec:paths}) of other ISDs. The TRC
server's address is in the TRC file of the ISD, allowing Alice to directly query
the server for other TRC files. In an extreme case where Alice does not trust
the provider or ISD, she may download a TRC file from a publicly-accessible
mirror site or obtain one in person from a trusted colleague or organization.
Alice can also obtain a TRC file \emph{a priori} if she plans to join such an
ISD with a new device.

\myparagraph{Obtaining updated TRC files} ASes and users in an ISD
are informed of the latest version of the TRC file with each routing
announcement and DNS response. Thus, as long as Alice has an Internet
connection and performs DNS lookups, she can quickly detect and obtain a
new TRC. The version number is part of each routing announcement, and a
timestamped message signed by the trust roots accompanies each DNS
\iflongpaper answer (to avoid re-signing every DNS record in the
ISD upon updating the TRC file).
\fi
When Alice detects a new TRC file, she can fetch the new file from the
provider or DNS
\iflongpaper root (see \autoref{fig:update}).
\fi

\myparagraph{Changing trust anchor ISDs} Besides the ability to select
trust roots as described in \autoref{sec:ead:anchor}, trust agility also
provides the ability to easily and quickly modify this selection. The above
methods of obtaining TRC files provide this notion of trust agility, as
Alice can change trust anchor ISDs by simply obtaining a new TRC file.
Under normal circumstances Alice can simply download a new TRC file from
her trust anchor ISD's TRC server, but if for example she discovers
that her trust anchor ISD has been conducting state-level surveillance, she
can instead obtain the TRC file manually or from an external server as
described above.

%% file: paths.tex
\ifnoindentaftersectionheadline\noindent\fi
\iflongpaper
In this section, we provide a detailed discussion of cross-signing in \name.
We begin by describing cross-signing certificates, and then detail how these are
used to enable inter-ISD authentication and global verifiability.
We then discuss the tradeoff between global verifiability and the
trustworthiness of authenticated information, and how the authentication
policies expressed in an ISD's TRC file fits in this tradeoff.
\fi

\subsection{Cross-Signing Certificates}

\noindent In order to enable global verifiability, we require ISDs to issue
\emph{cross-signing certificates} for its neighboring ISDs, that is, ISDs with
which they share routing links. The resulting web of cross-signing between ISDs
ensures that by following a route from Alice's ISD to Bob's ISD, a corresponding
chain of signatures from Alice's trust roots to Bob's trust roots will exist,
forming a chain of signatures from Alice's trust roots to Bob. ISDs without
direct routing connections can also issue cross-signing certificates to one
another, forming further chains of signatures to enable authentication between
different ISDs directly.

A cross-signing certificate issued by $\cI$ for $\cJ$'s trust roots contains
\begin{inparaenum}[(a)]
  \item a timestamp,
  \item $\cI$,
  \item the current version number of $T_\cI$,
  \item $\cJ$,
  \item the current version number of $T_\cJ$,
  \item a hash of $T_\cJ$, and
  \item a signature by a quorum of $\cI$'s trust roots
\end{inparaenum}
(see \autoref{tab:notation} for an explanation of notation). The version numbers
of the TRC files ensure that the trust roots' public keys can be checked
against the appropriate versions of the TRC files.

The ISD stores these certificates in its TRC server for its users and
also propagates the certificates along its inter-ISD routing links to
provide each ISD with the necessary information to form a chain of
signatures to a given destination ISD. Alice can then query her trust
anchor ISD to obtain these chains of signatures and select one to
authenticate information in Bob's ISD.

\subsection{Inter-ISD Authentication}

\noindent When Alice, whose trust anchor ISD is $\cK$, wants to
authenticates Bob, who is in another ISD $\cM$, she needs to obtain
cross-signing certificates to form a chain of signatures from $\cK$ to $\cM$.
While verifying Bob's routes, name, and EE certificate, she obtains the
appropriate cross-signing certificates from $\cK$'s TRC server. If Alice is
in an ISD $\cI$, then every route from her to Bob will have a chain of
signatures starting at $\cK$, proceeding to the trust roots of $\cI$, then to
the trust roots of $\cM$, and finally to Bob's AS.

If $\cK$ and $\cM$ do not share routing links but have issued cross-signing
certificates to each other, Alice can verify Bob's name and EE certificate
using $\cK$'s cross-signing certificate for $\cM$. These cross-signing
``shortcuts'' allow Alice to authenticate Bob's information with fewer ISDs
authenticating information ``in transit,'' providing fewer opportunities for a
compromised trust root to disrupt authentication.

No matter where $\cM$ is, Alice is guaranteed to find a chain of signatures to
$\cM$ and to Bob if she can find a route to Bob. Since she must be able to
contact $\cK$ from $\cI$ to obtain the appropriate cross-signing certificates,
she has a route between $\cK$ and $\cI$ and can thus obtain cross-signing
certificates from $\cK$ to $\cI$, and similarly for $\cI$ and $\cM$. Though a
chain of signatures may cross many ISDs, Alice is guaranteed to find at least
one such chain.

Note that cross-signing certificates do not necessarily indicate a trust
relationship between ISDs; a cross-signing certificate instead only states:
``These are the public keys of the trust roots for the following ISD.'' It is
therefore up to Alice to determine the trustworthiness of a chain of signatures
before accepting the information it certifies as authentic.

\subsection{Authentication Policies}

\noindent
The above cross-signing requirement ensures that Alice can authenticate Bob's
information regardless of which ISD he is in. While a compromised trust
root on a chain of signatures from Alice to Bob can adversely affect
authentication by certifying false information, Alice's trust anchor ISD
$\cK$ can mitigate this risk through the use of ISD-wide policies in the
TRC file. These policies can also blacklist public keys, such as those
contained in known unauthorized certificates or those of compromised trusted
authorities. Using such policies, $\cK$ can protect Alice from compromises in
other ISDs. If others with $\cK$ as their trust anchor ISD frequently
contact Bob or other destinations in $\cM$, then $\cK$ may form a cross-signing
relationship with $\cM$ to minimize the risk of compromised trust roots in other
ISDs.

ISDs face a tradeoff between enabling global verifiability and protecting their
users from compromises in other domains. The default behavior in \name is to
provide global verifiability. As illustrated above, an ISD must explicitly state
any exceptions to this behavior in the policy field of its TRC file. The ability
to restrict the authentication of known false information through policies
provides a mechanism by which an ISD can protect not only its own users, but
also users for whom a chain of signatures passes through the ISD.

%% file: separation.tex
\ifnoindentaftersectionheadline\noindent\fi
\iflongpaper
In this section, we describe how \name separates routing and service
authentication. We first describe our motivation for separating these two types
of authentication, and then discuss how this separation provides Alice with
trust mobility.
\fi

\subsection{Routing and Service Authentication}

\noindent Authentication in \name is classified and separated into routing and
service
\iflongpaper
authentication (see \autoref{fig:auth-triangle}).
\fi
We make
this separation in part because we observe that the authentication of route
information fundamentally differs from the authentication of service
information. In particular, routing authentication cannot assume the existence
of secure routes to obtain any external information, and therefore an entity
must rely on pre-verified paths or be able to verify paths without fetching
external information. By contrast, service authentication assumes the existence
of authentic routes and thus allows contacting external entities to obtain
authentication information.

Routing messages in \name propagate beginning from the ISD core and follow
provider-customer AS links. Unlike in RPKI and BGPSEC, all necessary information
(e.g. AS certificates) are sent with the routing message, allowing an AS to
verify routing messages upon arrival. Moreover, information such as AS
certificates are short-lived, eliminating the need to propagate revocation
information for AS keys.

By contrast, a DNS lookup, which falls under service authentication, must use a
route to reach one or more nameservers and fetch the appropriate information
for verifying a name-to-address mapping. TLS features such as
OCSP also require contacting an external entity to determine the validity of an
EE certificate.  Due to this dependence, Alice must verify routes to the ISD
core of her current ISD $\cI$, and form and verify routing paths from $\cI$ to
Bob's ISD $\cM$ before she can authenticate Bob's service information.

\subsection{Trust Mobility}

\noindent
Separating routing and service authentication also enables trust mobility.
Suppose that Alice checks into a hotel in Oceania, a known
surveillance state, and attempts to connect to her hotel's wireless Internet. If
the Oceanian trust roots are compromised by the government, then it is
inevitable that the government can see her packets themselves, as her physical
location in Oceania enables the government to examine her packets. In other
words, Oceanian trust roots must certify her routes out of the Oceanian
ISD and thus these trust roots must be on the chain of signatures for
routes from Alice to any destination in the Internet.

With \name, however, Alice can choose \cK as her trust anchor ISD for service
authentication, since \name separates routing and service authentication.
Moreover, this choice does not depend on Alice's current location and thus
applies wherever Alice is in the Internet. In our example, this means that Alice
does not have to rely on signatures from the Oceanian trust roots to verify
Bob's name or EE certificate, even if she is connecting to the Internet from an
Oceanian hotel.

%% file: validation.tex
\ifnoindentaftersectionheadline\noindent\fi We now discuss the complete
authentication process in \name. We first describe setup steps for a
\iflongpaper
server, such as joining an ISD and registering domain names, routing
paths, and EE certificates. We
\fi
then describe how Alice (the client) checks the information that she
receives about Bob (the server).
We use $a$ to denote Alice and $b$ to denote Bob. As previously mentioned,
Alice's trust anchor ISD is \cK. Bob is part of the AS $B$ in
Mythuania \cM, whose ccTLD is \texttt{.my}. \autoref{tab:notation}
provides a list of the notation used.

\begin{figure}[t]
  \scriptsize
  \begin{tabularx}{\linewidth}{|c|l|X|}
    \hline
    \!\!Notation\!\! & Name & Use\\
    \hline\hline
    \multicolumn{3}{|l|}{\textbf{Identifiers}} \\
    \hline
    $X$ & AS & AS with ASID $X$ \\
    $y$ & Endhost & an end-entity such as a client or server \\
    $e_y$ & \EID & locate endhost $y$ within its AS and ISD \\
    $\mathcal{Z}$ & ISD & ISD with identifier $\mathcal{Z}$ \\
    \hline
    \multicolumn{3}{|l|}{\textbf{Certificates}} \\
    \hline
    $\mathit{AC}_X$ & AS cert & bind $X$ to $AK_X$ (signed by $RS$ of $X$'s ISD) \\
    $\mathit{EC}_y$ & End-entity cert & store CA-signed public key information during connection setup \\
    $\mathit{DC}_y$ & CERT RR & store CA-signed DNS binding between $y$ and $DK_y$ \\
    \hline
    \multicolumn{3}{|l|}{\textbf{Keys}} \\
    \hline
    $AK_X$ & AS key & sign paths that can be used to reach $X$ \\
    $DK_y$ & DNSKEY RR & sign DNS resource records in DNSSEC \\
    $EK_y$ & End-entity key & set up secure end-to-end connections, e.g., via TLS \\
    $K_y^{-1}$ & Private key & private key for public key $K_y$ \\
    \hline
    \multicolumn{3}{|l|}{\textbf{Servers}} \\
    \hline
    $\mathit{PS}_Y$ & AS Path server & contact ISD path server for clients in $Y$ \\
    $\mathit{PS}_\cZ$ & ISD Path server & maintain database of signed paths for ASes in \cZ \\
    $RS_{\cZ}$ & Registry server & assign ASIDs and AS numbers in \cZ \\
    \hline
    \multicolumn{3}{|l|}{\textbf{Messages}} \\
    \hline
    $P_X$ & Signed path set & sent to $PS$ of $X$'s ISD to register paths to reach $X$ \\
    $T_{\cZ}$ & TRC file & provide trust root information for \cZ \\
    \hline
  \end{tabularx}
  \caption{Notation.}
  \label{tab:notation}
\end{figure}

\subsection{AS Setup}
\label{sec:validation:as-setup}

\begin{figure}[t]
  \centering
  \includegraphics[trim=0 20 0 0,width=0.7\linewidth]{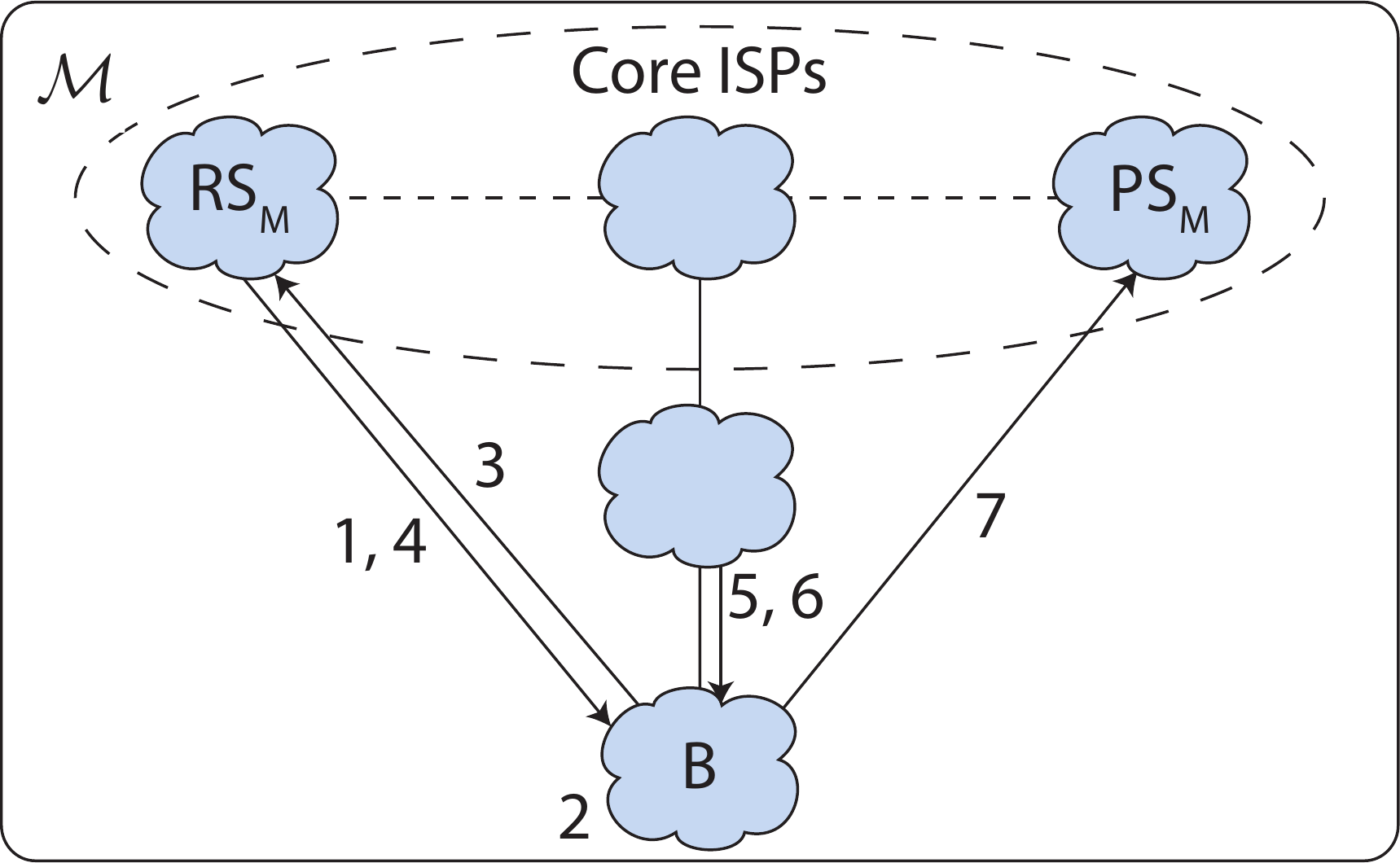}
  \caption{Diagram for AS setup steps (\autoref{sec:validation:as-setup}).}
  \label{fig:as-setup}
\end{figure}

\noindent
\autoref{fig:as-setup} depicts the steps of the AS setup process for AS $B$ in the
Mythuanian ISD \cM:
\begin{compactenum}

  \item $RS_\cM$ assigns the ASID $B$ to Bob's AS.

  \item $B$ creates an AS key pair $(AK_B, AK_B^{-1})$.

  \item $B$ sends $\left\{ B, AK_B \right\}$ to $RS_\cM$.

  \item $RS_\cM$ issues $B$ an AS certificate $\mathit{AC}_B$.

  \item $B$ receives the TRC file $T_\cM$ from its parent AS.

  \item $B$ receives SCION routing messages from its parent.

  \item $B$ selects a set of paths $P_B$ (signed with
    $AK_B^{-1}$) and sends $\left\{ P_B, \mathit{AC}_B \right\}$ to $\mathit{PS}_\cM$.

\end{compactenum}

\subsection{Server Setup}
\label{sec:validation:server-setup}

\begin{figure}[t]
  \centering
  \includegraphics[trim=0 15 0 0,width=0.75\linewidth]{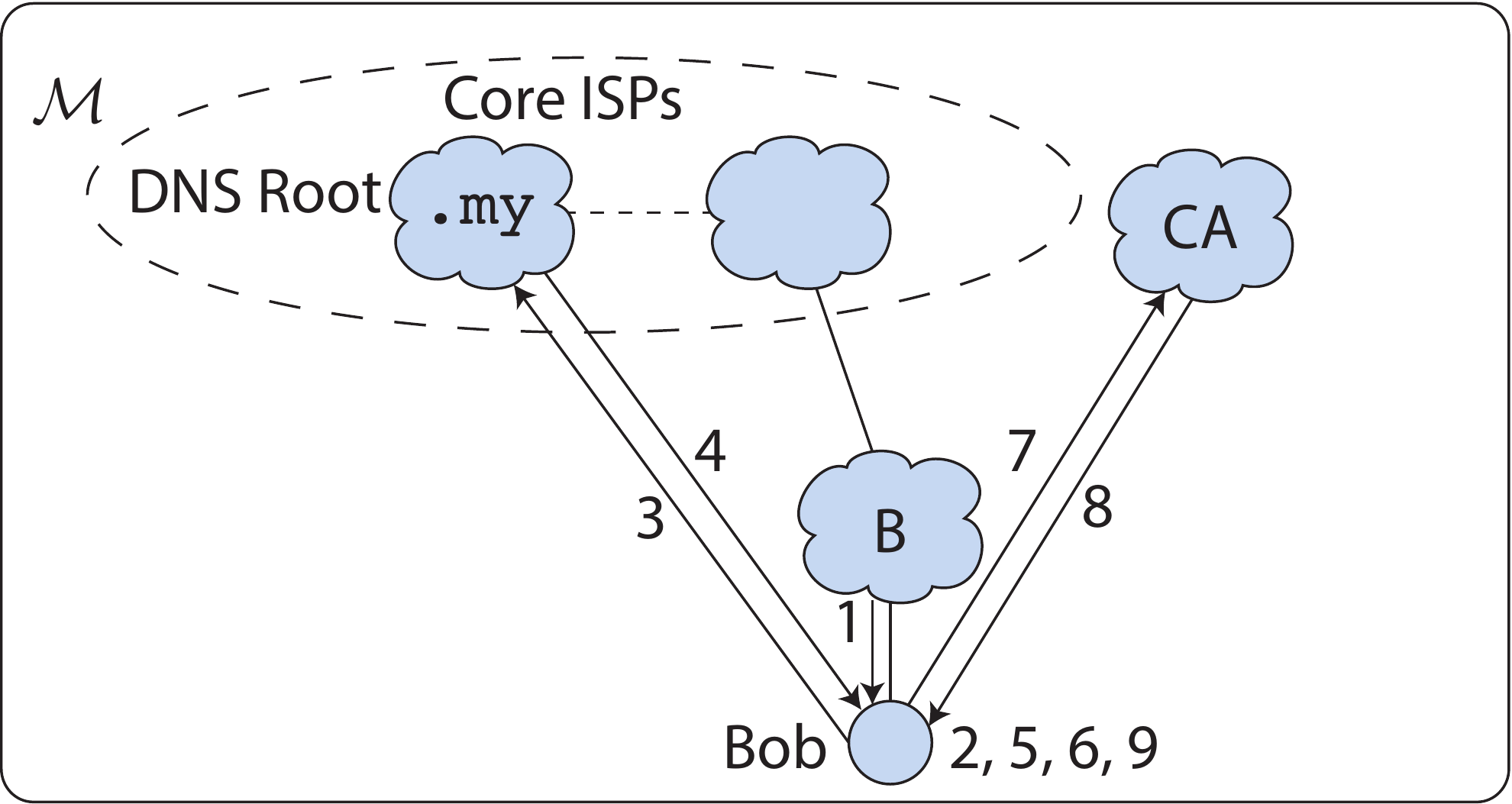}
  \caption{Diagram for server setup steps (\autoref{sec:validation:server-setup}).}
  \label{fig:server-setup}
\end{figure}

\noindent
\autoref{fig:server-setup} shows the steps of the server setup process for Bob:
\begin{compactenum}

  \item AS $B$ assigns Bob the \EID $e_b$, making his address $(\cM, B, e_b)$.

  \item Bob chooses the name \texttt{b.my} and creates a domain-name key pair
    $(DK_b, DK_b^{-1})$.

  \item Bob sends \texttt{b.my} and $DK_b$ to the \texttt{.my} operator to
    register his name and key.\footnote{In practice, Bob will create multiple key
    pairs and use one of the private keys to sign the others, but for simplicity
    we assume here that Bob uses $DK_b$ both to sign his DNS zone information
    and to self-sign $DK_b$.}

  \item The \texttt{.my} operator creates a delegation signer (DS) record to
    point to $DK_b$ from the \texttt{.my} zone, as well as a record mapping
    \texttt{b.my} to Bob's nameserver.

  \item Bob creates a mapping of \texttt{www.b.my} to $(\cM, B, e_b)$,
    $DK_b$, and resource record signature (RRSIG) record of the mapping signed with $DK_b^{-1}$.

  \item Bob creates an end-entity key pair $(EK_b, EK_b^{-1})$.

  \item Bob sends $\left\{ b, EK_b \right\}$ to a CA in \cM.

  \item The CA issues Bob an EE certificate $\mathit{EC}_b = \left\{ b, EK_b
    \right\}_{K_{CA}^{-1}}$.

  \item Bob creates a certificate $\mathit{DC}_b = \left\{ b.my, DK_b
    \right\}_{EK_b^{-1}}$, and stores $\mathit{DC}_b$ along with $\mathit{EC}_b$ as a CERT record
    in his nameserver.

\end{compactenum}

\subsection{Client Setup}
\label{sec:validation:client-setup}

\begin{figure}[t]
  \centering
  \includegraphics[trim=0 15 0 0, width=0.75\linewidth]{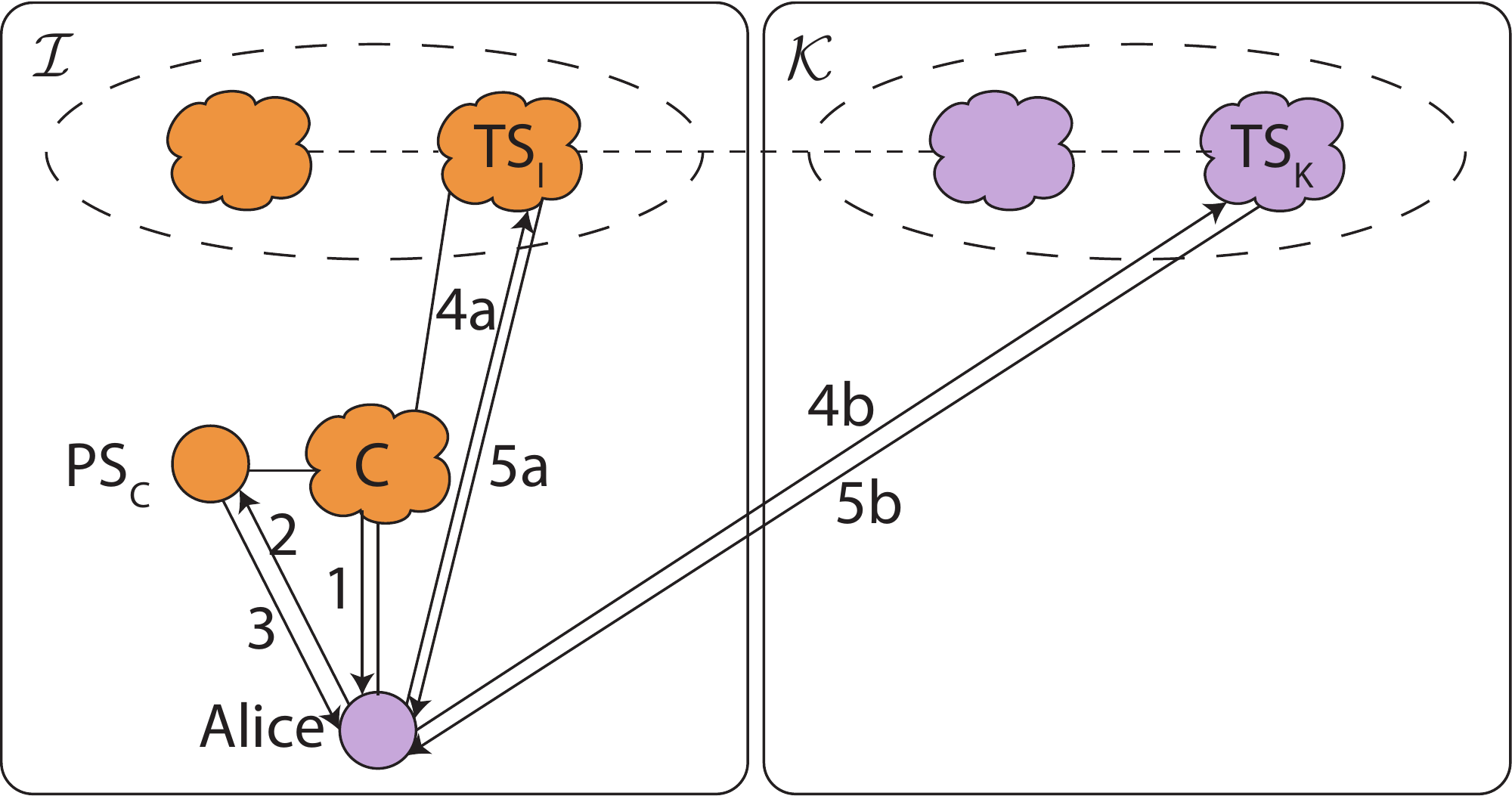}
  \caption{Diagram for obtaining a TRC file (\autoref{sec:validation:client-setup}).}
  \label{fig:client-setup}
\end{figure}

\noindent
In order for Alice to verify information, she must first possess a TRC file
to configure her set of trust roots. Even if she does not have a TRC file,
we assume that she can verify a TRC file from her trust anchor ISD
\cK. In fact, she can verify this TRC file in any ISD \cI --- even
if she does not trust \cI. As illustrated in \autoref{fig:client-setup}, after
connecting to the Internet in \cI, Alice does the following to obtain
and verify a TRC file:

\begin{compactenum}

  \item Alice's ISP (AS $C$) assigns her the \EID $e_a$, making her
    address $(\cI, C, e_a)$. $C$ also sends her the latest TRC file
    $T_\cI$ for the ISD \cI.

  \item Alice requests from $\mathit{PS}_C$ a path to the TRC server $\mathit{TS}_\cI$ of \cI
    or $\mathit{TS}_\cK$ of \cK (if she knows the address).

  \item $\mathit{PS}_C$ returns to Alice the path she requested.

  \item Alice now contacts either $\mathit{TS}_\cI$ (4a in \autoref{fig:client-setup}) or
    $\mathit{TS}_\cK$ (4b) and requests $T_\cK$. In the case of 4b, Alice also requests a
    cross-signing certificate for \cI from $\mathit{TS}_\cK$ to ensure that all
    authentication (even for routes) begins from $T_\cK$.

  \item If Alice contacted $\mathit{TS}_\cI$, she receives $T_\cK$ (5a). Otherwise, she
    receives $T_\cK$ and the cross-signing certificate for \cI from $\mathit{TS}_\cK$
    (5b).

\end{compactenum}

We assume that Alice verifies the authenticity of $T_\cK$ through an out-of-band
mechanism, e.g., if she makes plans to travel to \cI and considers it a
``hostile'' ISD, then Alice can obtain a hash of the public keys of \cK's
trust roots \emph{a priori}, or she can obtain this information in an embassy of
\cK within \cI.

\subsection{Client Verification}
\label{sec:validation:verification}

\begin{figure}[t]
  \centering
  \includegraphics[trim=0 15 0 0,width=\linewidth]{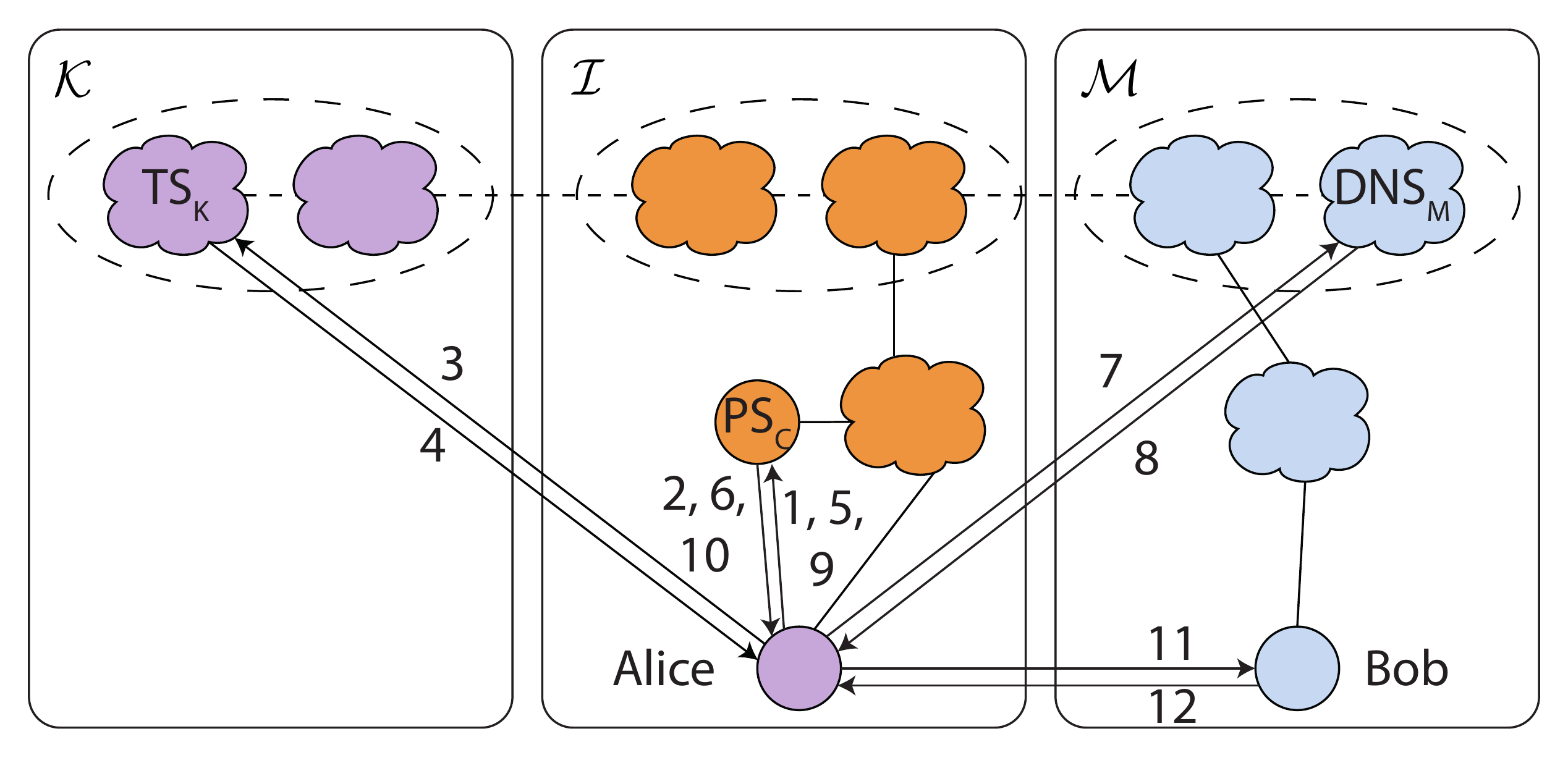}
  \caption{Client lookup and verification of server's name, route, and EE
    certificate (\autoref{sec:validation:verification}).}
  \label{fig:verification}
\end{figure}

\noindent
\autoref{fig:verification} illustrates the complete process that Alice executes
to authenticate Bob. We assume that Alice has completed the client setup process
and thus has $T_\cK$ to use as the starting point for authenticating Bob. The
authentication process is as follows:

\begin{compactenum}
  \item Alice begins by looking up \texttt{www.b.my}, and thus first
    obtains $T_\cM$. She contacts $\mathit{PS}_C$ to obtain a path to $\mathit{TS}_\cK$.
  \item $\mathit{PS}_C$ returns to Alice a set of paths that she can use to reach
    $\mathit{TS}_\cK$.
  \item Alice contacts $\mathit{TS}_\cK$ to request $T_\cM$.
  \item $\mathit{TS}_\cK$ returns $T_\cM$ and a cross-signing certificate for
    \cM.
  \item Alice contacts $\mathit{PS}_C$ to request a path to the DNS root of \cM, whose
    address she has from $T_\cM$.
  \item $\mathit{PS}_C$ returns to Alice a set of paths to \cM's DNS root.
  \item Alice contacts \cM's DNS root to query \texttt{www.b.my}.
  \item Alice performs DNSSEC resolution to obtain $(\cM, B, e_b)$ as well
    as Bob's domain and EE certs $\mathit{DC}_b$ and $\mathit{EC}_b$.
  \item Alice requests a path to Bob's address from $\mathit{PS}_C$.
  \item $\mathit{PS}_C$ returns to Alice $B$'s AS certificate $\mathit{AC}_B$ and a set of paths
    $P_B$ to reach $B$.
  \item Alice contacts Bob to initiate the TLS handshake.
  \item Bob sends Alice his EE certificate $\mathit{EC}_b$.
\end{compactenum}

Alice verifies that Bob's EE public key $EK_b$ contained in the EE certificate
she obtained from \cM's DNS root matches the EE public key she receives during
the TLS handshake. If the keys match, she proceeds with the TLS handshake to
establish a secure end-to-end connection with Bob.

Throughout this process, Alice verifies that valid authentication paths exist
for each entity she contacts: $\mathit{PS}_C$, $\mathit{TS}_\cK$, \cM's DNS root, $B$, and Bob.
When she receives information signed by the trust roots of an ISD other
than \cK, Alice uses the appropriate cross-signing certificate to verify the
public keys of the ISD's trust roots, thus ensuring that all authentication
ultimately begins with trust roots listed in the TRC of her trust anchor
ISD \cK.

\iflongpaper
\myparagraph{Error handling} A verification failure at any stage in the
authentication process will prevent Alice from authenticating and establishing a
connection to Bob. In the event that the verification of a routing path fails,
Alice will not be able to reach Bob or entities such as DNS roots and TRC
servers. However, Alice likely cannot detect this failure from her browser. In
the event that the verification of Bob's name-to-address mapping fails, Alice
will not know the address at which she can reach Bob. While most modern browsers
indicate such a failure, Alice cannot proceed with verification after such a
failure. From the perspective of Alice's browser, a failure to verify Bob's EE
certificate is the most informative, as most modern browsers display the type of
error that occurred and in some cases provide the option to continue with the
connection anyway.
\fi

%% file: evaluation.tex
\iflongpaper
\ifnoindentaftersectionheadline\noindent\fi
In this section, we describe our prototype implementation of \name. We used
this implementation to evaluate the performance of authentication and trust root
management functions; we also discuss our evaluation results here.
\fi

\subsection{Implementation}

\noindent
We implemented the endhost side of \name\iflongpaper (see
\autoref{fig:endhost_impl})\fi. The main component of our implementation is the
\emph{\name daemon}, which acts as a gateway between applications and the
network. The \name daemon includes SCION layer support for packet
encapsulation and decapsulation, a path engine for route management and
verification, a name lookup  engine for \name name queries, and a
TRC engine, which allows users to obtain and verify TRC files.

\iflongpaper
\begin{figure}[t]
  \centering
  \includegraphics[width=0.8\linewidth]{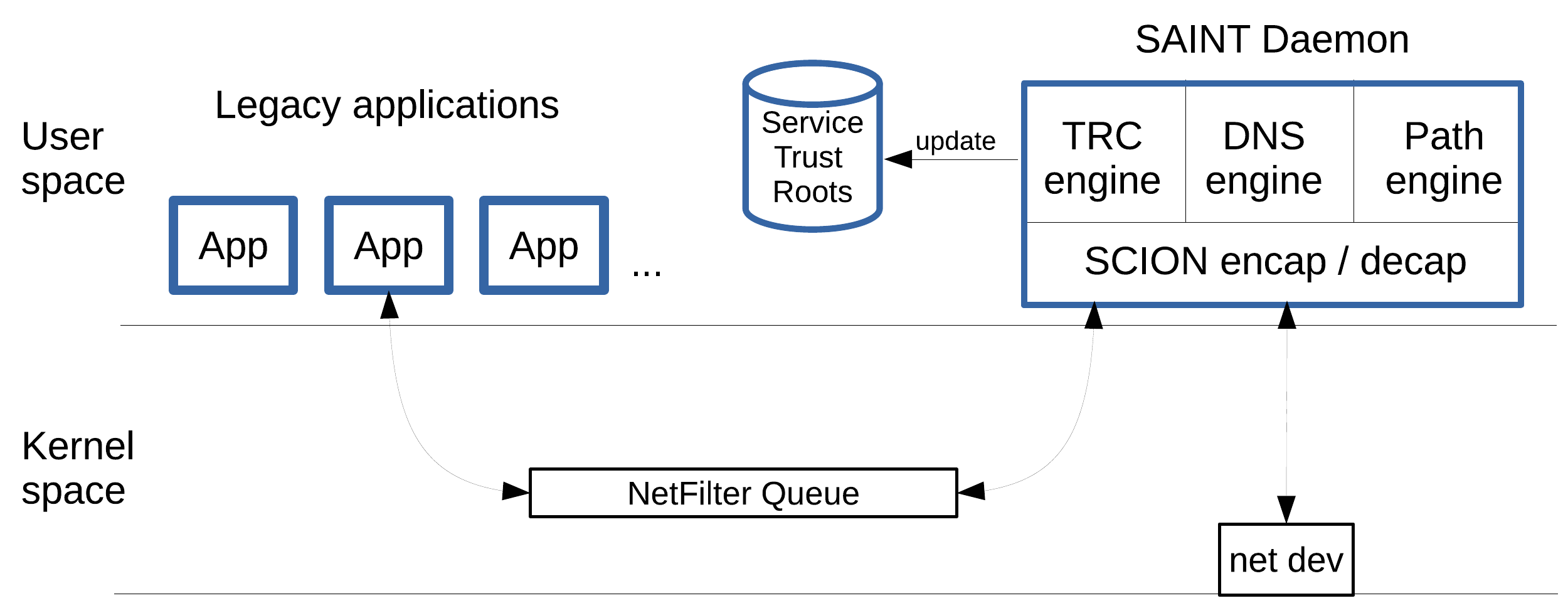}
  \caption{Architecture for endhost implementation.}
  \label{fig:endhost_impl}
\end{figure}
\fi

Traffic generated by the applications is delivered to the \name daemon by
the NetFilter queue, allowing legacy applications to deploy \name without
requiring any changes.
We
ran our simulation on an Intel Core i5-3380M CPU at 2.90
GHz, 16 GB of RAM, Python 3.4, and gcc 4.8.2. We used
ed25519~\cite{bernstein2012high} as our signature scheme for name and path
verification, and RSA-2048 for TLS certificates.

\subsection{TRC Updates}
\label{sec:evaluation:trc}

\begin{figure}[t]
  \centering
  \includegraphics[trim=0 15 0 0,width=0.87\linewidth]{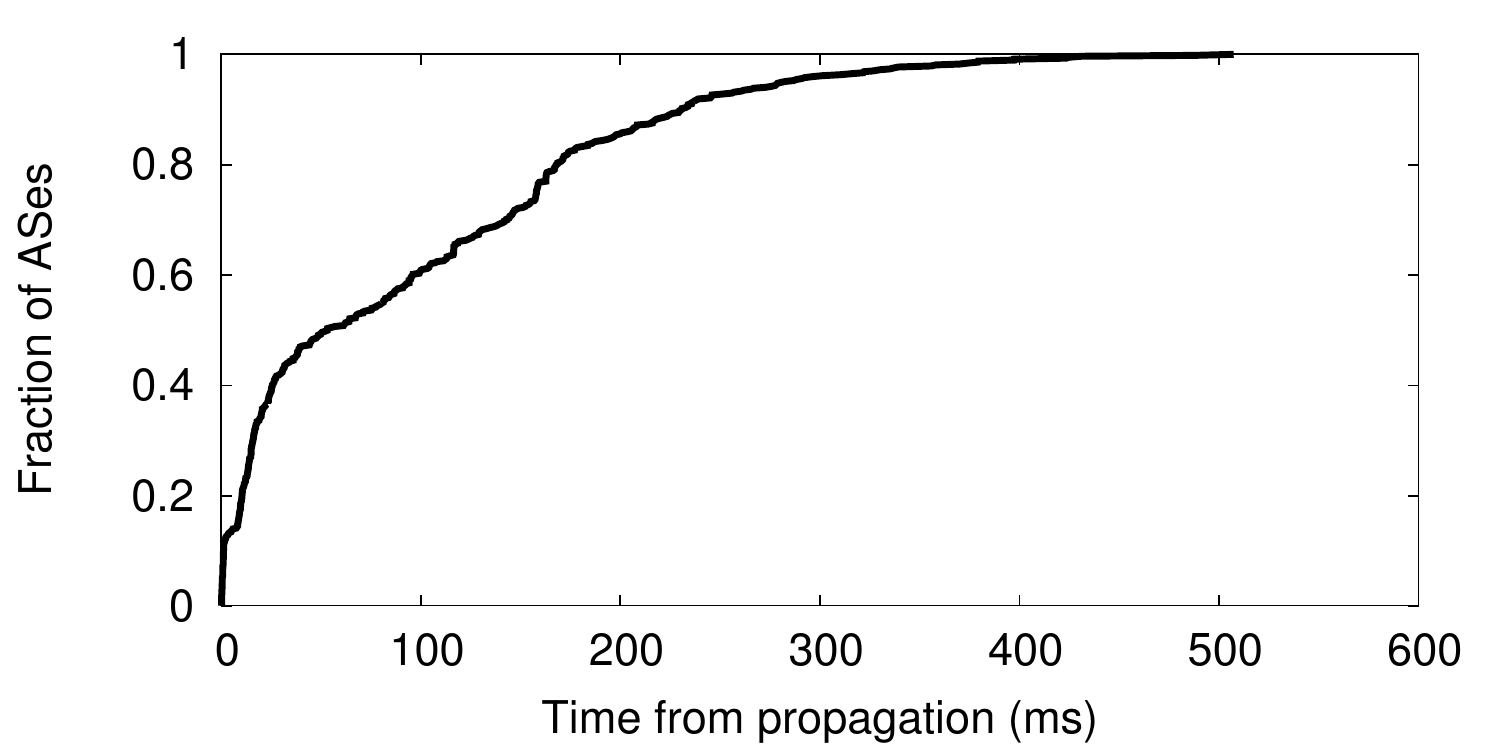}
  \caption{CDF of the percentage of vantage point ASes using a new TRC after
  propagation from the core ISPs.}
  \label{fig:new-trc}
\end{figure}

\noindent
We measured the efficiency of TRC distribution and updates by simulating
the propagation through the current AS topology. We used the CAIDA inferred AS
relationships dataset from October 2014~\cite{as-rel} as our model of the
current topology, and used traceroutes from iPlane
datasets\footnote{http://iplane.cs.washington.edu/data/data.html} to estimate
inter-AS latency. From each of iPlane's vantage point ASes (distributed
throughout the world), we identified the latency (half of the RTT) to each of
the top-tier Ases identified using the CAIDA-based topology.

Our evaluation demonstrates that more than half of our vantage point
ASes receive a new TRC file within 100 ms of the file being sent
from the core
\iflongpaper ISPs (see \autoref{fig:new-trc}).
\fi
Moreover, all vantage point ASes receive the new TRC file
within 600 ms. Our vantage point ASes included stub ASes (that is,
ASes with no customer ASes), demonstrating that end users around the
world can quickly receive updated TRC files. These results show
that our TRC propagation mechanism is significantly more
efficient than the current trust root update mechanisms in browsers
(which typically occur on the order of days).

\subsection{Authentication Overhead and Performance}

\noindent
To test the actual latency of authentication in \name, we measured 300
end-to-end secure connection establishments on a sample SCION topology of
virtual machines. \autoref{tab:eval} shows the timing results, which
only reflect the cryptographic verifications and do not take into
account network delay,
since a full network deployment of SCION is not available at this time. However,
our results give us an insight into the overhead of \name. In particular, all
functions take less than 1 ms on average, which is significantly less than the
end-to-end round trip time in an actual connection establishment.

\begin{figure}[bt]
  \scriptsize
\begin{center}
\begin{tabular}{|r|r|r|r|r|}\hline
Measurement & Min & Max & Med & Avg \\\hline \hline 
DNS resolution & 199 & 967 & 557 & 500 \\ \hline
Path verification & 348 & 1\,691 & 691 & 652 \\ \hline
Certificate validation (TLS) & 210 & 1\,123 & 222 & 233 \\ \hline
Certificate validation (TLS+CRL) & 401 & 1\,775 & 440 & 460 \\ \hline
\end{tabular}
\end{center}
\vspace{-5mm}
\caption{Evaluation results (in microseconds).}
\label{tab:eval}
\end{figure}

As a baseline, we measured end-to-end connection establishments to 100 HTTPS
sites on the current Internet randomly selected from \texttt{httpsnow.org}.
We observed 418 separate TLS connection
establishments. Using DNSSEC, BGPSEC, and TLS,
 we measured the latency of the total page loading time, which included
blocking of the connection request, the DNS lookup, the connect, send, and wait,
and receive times, and the TLS handshake.  The observed latencies ranged from 15
ms to 7\,969 ms, with an average of 534 ms and a median of 1\,811 ms.
For \name, we assumed that a path lookup has an equivalent
latency to a DNS query (one round trip), fast cryptography is used in
verification (ed25519), and an average of 8 keys are authenticated during
the DNS and routing lookups. In this
setting, we observed
latencies ranging from 19 ms to 7\,974 ms, with an average of 586 ms and a median
of 1\,863 ms. These latencies are not based on a full deployment of
\name, but indicate a reasonable 10\% increase in connection latency on average as compared
to the current Internet.

%% file: deployment.tex
\iflongpaper
\ifnoindentaftersectionheadline\noindent\fi
We now discuss several deployment challenges for \name and propose several
solutions to facilitate the deployment of \name in the current Internet.
Specifically, we discuss \name's interoperability with the current Internet,
and describe how ISDs can be initially deployed. We then propose a method
for the initial distribution of TRC files.
\fi

\myparagraph{The Legacy \ead}
To facilitate the incremental deployment of \name, we propose a special
ISD called the \emph{legacy ISD} \cL, which represents the set of all
domains that have not yet joined a \name ISD. For example, suppose that
\domain{a.com} maps to the IP address 1.2.3.4 in the current Internet, which is
located in AS 567, which has not yet deployed \name. The domain
\domain{a.com} would then correspond to the address $(\mathcal{L}, 567,
1.2.3.4)$. The security guarantees for names, routes, and EE certificates in the
legacy ISD are only as strong as they are in today's Internet.

\iflongpaper
One challenge we face in practice is that names in \name's generic TLDs may
already exist in the legacy \ead. Not only do such collisions cause a problem
for name resolution, but they also create vulnerabilities to downgrade attacks
in \name's name lookup mechanism, since an adversary could simply return an
unsecured DNS response in the legacy \ead for a query with a name collision. We
therefore require \name name resolvers to query the legacy \ead as a last
resort, and only with a proof of the name's absence in the \name namespace (such
as an NSEC3 record).
\fi

\myparagraph{Interoperability}
\iflongpaper
\name is designed with a focus on incremental deployability.
\eads can be deployed among individual networks, since the remainder
of the Internet that has not yet deployed \name joins the
the legacy ISD.
The naming infrastructure is interoperable with that of the current
Internet in that it can query the legacy DNS nameservers as is done
today.
Routing between two physically separated domains in SCION can utilize
IP tunneling to communicate over the legacy \ead.

\fi
In order to maintain connectivity to the current Internet, servers and clients
must support legacy authentication. In particular, clients and servers must
continue to support DNSSEC, BGPSEC, and TLS. Additionally, when servers
receive incoming connections from the legacy ISD, they should not respond
with \name-specific messages such as signed path sets or cross-signing
certificates. However, TRC files will be made
available through the legacy ISD in order to support the initial
bootstrapping of trust roots.

\myparagraph{ISD Deployment} \name offers benefits even for early
deployers of ISDs. A single-\ead deployment of \name provides
isolation of compromises within that \ead. Additionally, the legacy
ISD will be protected from compromised trust roots in every ISD
deploying \name. Moreover, ISDs deploying \name enjoy greater
flexibility in their choice of alternative PKIs by enabling the
benefits of the PKIs without requiring their global deployment. In
the policy field of the TRC file, an ISD would specify its choice of
the underlying PKI, which would prevent protocol downgrade attacks.

If using countries as \eads, then a newly-deploying \ead can simply
attach to the existing namespace at its corresponding ccTLD. This
construction allows the DNS to provide a scaffolding during the
deployment of \name and also allows the DNS in \name to distribute
TRC files.

\iflongpaper
However, we recognize the challenges that come with using countries as
ISDs. In particular, the deployment of such a scheme would require the core
ISPs, root CAs, and Internet registries in each country to create a federation
of trust roots. In practice, we may see corporations rather than countries form
ISDs. In this case, ISDs would have to form IP tunnels in order to form
inter-ISD routing relationships. Additionally, since there are far more
corporations than countries, cross-signing relationships may not scale to this
number of ISDs. However, because most corporations do not form many
business relationships relative to the number of corporations that exist, we do
not expect that the number of cross-signing relationships will grow to an
unsustainable scale.
\fi

\myparagraph{TRC Distribution}
\iflongpaper
The initial distribution of TRCs must occur securely since TRCs are the starting
point for all authentication in \name. Many trust roots in the current
Internet may continue to serve as trust roots in \name, and thus may be able to
``inherit'' user trust in \name that they already have in the current Internet.
However, \name will likely result in the creation of new trust roots, and thus
must have a mechanism for bootstrapping trust in the initial public keys of
these roots.

To address this challenge,
\fi
we suggest to perform the initial distribution of
\name TRC files through \dnssec. Since ISDs can deploy by attaching to
specific ccTLDs in the current DNS namespace, an ISD can create a reserved
domain name such as \domain{trc.us}, whose DNS record contains the TRC (e.g., in
a TXT record). Clients can then fetch the TRC by looking up the appropriate
domain name. 
\iflongpaper
Additional work has been done in distributing authentication
information through out-of-band means such as over public
radio~\cite{schulman2014revcert}, but these strategies are beyond the
scope of this paper.
\fi

%% file: discussion.tex
\myparagraph{Feasibility of country-based \eads}
In order to determine the feasibility of having countries as \eads in \name as
described in \autoref{sec:infrastructure:isd}, we mapped AS numbers to countries
and examined the resulting inter-\ead relationships. We used the AS
relationships database from CAIDA~\cite{as-rel} and Team Cymru's IP to AS number
mapping tool,\footnote{http://www.team-cymru.org/Services/ip-to-asn.html} to map
AS numbers to countries. We identified 228 ``countries'' in total, including the
EU and ZZ (indicating that the AS's country was unknown). We identified 2\,636
unique country pairs between which an inter-AS link 
 existed. These links signify direct routing connections, and thus we
expect cross-signing for each ISD pair. The
most prolific cross-signing ISDs were the US (196), the EU (135), and the UK
(124), but half of the ISDs cross-sign on the order of tens of other ISDs.

\myparagraph{Political Concerns} Given concerns over governmental nation-wide
surveillance, readers may worry about centralizing trust roots in a large \ead.
While we acknowledge that states may compromise these entities on a large scale,
\name's trust agility allows users to protect themselves from the interception
of sensitive connections.  Additionally, our efficient method of updating trust
root information allows \eads to quickly recover from a compromised CA.

Another concern we anticipate is that \name encourages fragmentation in the
Internet. While this is true, \name is designed to preserve global
reachability while simultaneously protecting users through the use of
ISDs and trust agility. We thus structure inter-ISD authentication
to provide users with the best of both worlds.

\iflongpaper
\myparagraph{Sub-\eads}
To add further scalability to \name, we propose the use of
\emph{sub-ISDs}, i.e., ISDs being completely contained in other
ISDs. A sub-\ead can be used to provide additional policies and
finer-grained control of trust roots in an ISD, and can additionally enable
isolated authentication within an \ead. For example, governments or medical
organizations may use their own network within a country \ead to ensure data
privacy and further scope the authority of their trust roots.

A sub-\ead structurally resembles a top-level \ead, but
authenticates to other sub-\eads in the
same parent \ead using the core of the parent \ead, and has its trust roots of a
certified by its parent
\ead via a cross-signing certificate.  Connections within
an \ead could then be negotiated using the lowest-level common \ead. For
example, two hospitals in an ISD would share data about a patient using the
medical sub-\ead rather than the general \ead.

Some details regarding authentication for sub-ISDs still remain, however.
For example, compromised trust roots could also affect authentication entirely
within a sub-\ead due to the requirement that parent ISDs certify the trust
roots of their sub-ISDs. Furthermore, users cannot select sub-ISDs as
their trust anchor ISDs without also trusting the corresponding parent
ISDs. We hope to investigate these challenges in future work.
\fi

\iflongpaper
\myparagraph{Optimizations}
As described in \autoref{sec:validation}, the authentication process involves
six round-trip connections from Alice. Each communication with the path server
requires verification of the path server's signature, the signed set of routing
paths returned by the path server, the destination AS certificate, and possibly
cross-signing certificates for the path server and destination AS's ISDs.
To contact a destination outside her trust anchor ISD, Alice must also
obtain and verify a cross-signing certificate. Contacting the DNS server
requires verification of at least the DNS root key, server DNS key, and signed
DNS record, and contacting the server (Bob) requires verification of at least
the server certificate.

However, in practice many of these verifications may not be necessary. Caching
cross-signing certificates, for example, eliminates the need to reach a TRC
server and to verify a cross-signing certificate with each end-to-end connection
establishment. Additionally, some of these verifications can be handled by
entities other than the client; for example, paths can be verified by the
client's AS, and DNS records by a trusted DNS stub resolver. Since ASes and stub
resolvers serve multiple clients, caching verification results can further
reduce the connection latency, especially for popular names, routes, and EE
certificates.

In order to further decrease the size of messages sent in the network, we can
also split the TRC file into routing and a service TRC files. The
routing TRC can then be propagated along AS links, and the service
TRC can be obtained from the TRC server. This scheme ensures that
users only receive the portion of the TRC that they need for a particular
type of authentication, thus reducing the size of TRC files sent in the
network.
\fi

%% file: related.tex
\iflongpaper
\ifnoindentaftersectionheadline\noindent\fi
This section provides related work supplemental to the work mentioned
in \autoref{sec:background}. Many such works propose mechanisms to
authenticate network entities. We review these works with regards to
domain-centric proposals as well as authentication for naming, routing, and EE
certification.
\fi

\myparagraph{Domain-centric proposals} The idea of aggregating hosts and routers
into an abstracted routing entity has been previously proposed. The Nimrod
routing architecture~\cite{rfc1992} describes a hierarchy of ``clusters'' of
hosts, routers, or networks that can reach each other via a path contained
within the cluster. FARA~\cite{ClBrFaPi2003} generalizes the notion of an
``entity'' to also include clusters of computers that can be reached as a
network communication endpoint. ISDs in \name fit the criteria for clusters in
Nimrod, but add a set of common trust roots as well as the constraint that
\emph{all} intra-domain paths must be contained within the ISD.

\myparagraph{Name authentication} Previous work has addressed authentication in
a distributed, large scale network without any global trust infrastructure.
Birrell et al.~\cite{BiLaNeSc86} propose to use an authenticated path through
the name space to make explicit trust relationships among entities, and Lampson
et al.~\cite{LaAbBuWo91} describe an authentication theory based on the name
space or the communication channel from which the other entity's authority can
be deduced. Gligor et al.~\cite{GlLuPa92} define a policy for inter-realm
authentication trust based on trust hierarchies that can support transparent
name authentication.

\myparagraph{Routing authentication}
AIP~\cite{AIP} provides accountability for
network entities based on self-certifying names, where the name of an
entity is its public
key~\cite{HIP-arch,HIP,MaKaKaWi99,WaStKrBaMoSh04}.
AIP groups an independent administrative network into an
accountability domain (AD) and assigned globally unique self-certifying names to
ADs and hosts.
\iflongpaper
Consequently, at the AD granularity, AIP not only
supports routing and forwarding authentication, but also domain
authentication without relying on an external PKI.
\fi
However, key
discovery in AIP relies on DNSSEC, and key revocations always force entities to change their names.

IPA~\cite{li2011bootstrapping} focuses on incremental
deployment in the current Internet and
leverages DNSSEC as a lightweight PKI to enable host
authentication. IPA distributes AS certificates via S-BGP routing update
messages, avoiding circular dependencies. However, it relies in a single global
root of trust.

\myparagraph{End-entity authentication} Several proposals raise issues with the
current domain authentication schemes based on X.509 and propose enhancements.
For example, CAge~\cite{kasten2013cage} proposes to restrict CAs to signing
domains in a small number of TLDs and treat other certificates as suspicious,
and the US government has also recently considered this
proposal~\cite{ca-letter}. Abadi et al.~suggest a policy engine to empower
clients or ISPs to specify acceptance criteria for
certificates~\cite{AbBiMiWoXi13}.
\iflongpaper
Although the authors outline a promising
user-oriented entity authentication policy, its integration in the end-user
system is still in question.
\fi

DANE~\cite{rfc6698} leverages the DNSSEC infrastructure to
authenticate TLS public keys.
\iflongpaper
Its goals are to tie TLS public keys to DNS names,
use DNS to distribute these public keys, and to leverage the hierarchical
authentication structure of DNSSEC to restrict the scope of CAs' authority.
\fi
However, the security of DANE relies on the security of DNSSEC.
A compromised DNSSEC key can be used to specify
arbitrary trust anchors and bypass X.509 certificate validation.

Certificate Transparency (CT)~\cite{laurie2013certificate} and the Accountable
Key Infrastructure (AKI)~\cite{kim2013accountable} expose all CA operations to
the public to improve the security of SSL/TLS PKIs. Neither CT nor AKI
define how misbehavior should be disseminated to users and other parties, but
both can be deployed in \eads, which leverage TRC files to quickly remove
a compromised trust root and update users' trust root information.

%% file: conclusion.tex
\ifnoindentaftersectionheadline\noindent\fi By explicitly separating and scoping
trusted authorities in the Internet, we allow users to choose their
trust roots and protect users from compromises throughout the Internet. By
distributing trust root information as network messages, we allow users to
quickly obtain up-to-date information about compromised or updated trust roots.
By mandating cross-signing relationships based on routing connections, we ensure
that users can authenticate information throughout the Internet. By separating
routing and service authentication, we allow users' trust root decisions to
apply anywhere in the world. These ideas address fundamental shortcomings of
current authentication and secure Alice's communications throughout the world
regardless of her choice of trust roots.

%% file: acknowledgments.tex
\ifnoindentaftersectionheadline\noindent\fi
We would like to thank Yih-Chun Hu, Virgil Gligor, Ari Juels, Burt Kaliski, and
Gene Tsudik for their insightful comments and guidance on drafts of this paper.
\iflongpaper
The research leading to these results has received funding from the
European Research Council under the European Union's Seventh Framework
Programme (FP7/2007-2013) / ERC grant agreement 617605.

We also gratefully acknowledge support from the NSF under grants CNS-1040801 and
DGE-1252522, from ETH Zurich, and from Google.
\fi

%% file: tec-dns.tex
\subsection*{DNS Resolution for Generic TLDs}
\label{sec:tec:dns:gTLDs}

\noindent
As stated in \autoref{sec:ead:namespace}, for reasons of transparency and
security, each domain name with a generic TLD is resolved to a regional domain
name (rather than to an address). Assume, for example, a company $r$ wants to
register the domain \domain{r.com}.  Instead of registering a DNS tuple
$(\domain{r.com},\mathit{IP})$, the company registers one or more CNAME-like
records\footnote{CNAME records cannot point to multiple names, but the general
idea of our records is the same.} that point to other (usually regional) domain
names:
\begin{equation} \domain{r.com} \;\to\; \{ \domain{r.us}, \;\; \domain{r.de},
  \;\; \domain{r-swiss.ch}, \;\; \domain{r-italia.it} \} \label{eq:dns}
\end{equation}

Upon a DNS resolution request for a generic domain $D$ from a
client, the DNS server for \domain{.com} returns all regional names
for $D$ in the order specified by the registrant $r$. The
client then either chooses a domain name that is within its own ISD, or
it chooses any other domain name in the provided list.

In order to ensure the authenticity of generic DNS records, \name
requires a minimal setup as follows: any registrant $r$ must first
register its DNS public key $DK_r$ with the generic DNS server $S$.
\begin{equation*}
  r \xrightarrow{\qquad\quad DK_r\qquad\quad} S \;\domain{(.com)}
\end{equation*}
$S$ stores and signs the public key $DK_r$, and returns the
signature $\left\{ r, DK_r \right\}_{K_S^{-1}}$.
\begin{equation*}
  r \xleftarrow{\qquad\left\{ r, DK_r \right\}_{K_S^{-1}}\qquad} S \;\domain{(.com)}
\end{equation*}
After this initial step, registrant $r$ registers its
domain name $D$ by providing the list of regional domain names $\{...\}$ together
with a signature $\left\{ D, \left\{ \ldots \right\} \right\}_{DK_r^{-1}}$.
\begin{equation*}
  r \xrightarrow{\qquad\qquad  \left\{ D, \left\{ \ldots \right\} \right\}_{DK_r^{-1}}   \qquad\qquad} S\;\domain{(.com)}
\end{equation*}

\smallskip
\myparagraph{Verification}
This domain-specific signature is verified whenever a client~$c$
wants to authenticate a DNS response for a generic domain~$D$.
\begin{equation*}
  c \xrightarrow{\qquad\qquad\qquad\qquad    \;\;\; D? \;\;\;    \qquad\qquad\qquad\qquad} S\;\domain{(.com)}
\end{equation*}
\begin{equation*}
  \xleftarrow{\quad  \left\{ D, \left\{ \ldots \right\} \right\} \quad
  \left\{ D, \left\{ \ldots \right\} \right\}_{DK_r^{-1}}\quad DK_r \quad
  \left\{ r, DK_r \right\}_{K_S^{-1}}    \quad} \qquad\qquad
\end{equation*}
The client verifies the signature $\left\{ r, DK_r \right\}_{K_S^{-1}}$ and
caches the public verification key $DK_r$ of the registrant~$r$ of
domain~$D$. The client then verifies the authenticity of the record
$\left\{ D, \left\{ \ldots \right\} \right\}$ using the registrant's public key $DK_r$.

If the verification was successful, the client choses one of the
specified regional domain names and resolves its actual address
$(\cI,A,E)$.

\smallskip \myparagraph{Performance} As in the current DNSSEC, the DNS
server $S$ for generic TLDs does not sign the stored CNAME-like records itself;
rather, it signs the public keys of the registrants. The reasons include
performance considerations: whenever a registrant $r$ wants to register a new
generic domain or whenever $r$ wants to extend an existing record, the DNS
server has minimal effort in that it does not need to validate or sign the new
records. Using CNAME-like records also keeps the performance close to existing
lookups: CNAME records add only 13\% latency to a DNS name resolution on
average.\footnote{This result is based on our private discussions with Verisign
Labs.}

\smallskip \myparagraph{Availability} Whenever a client needs to
resolve a generic domain name, the client first contacts its local
DNS server. If the local DNS server has no cached entry for the
generic domain, the request is redirected to the DNS server of
the generic TLD. This one step of
indirection is at least as robust as today's DNS system: in case the
DNS server for a generic domain is unavailable, then only the
availability of that single TLD is constricted. There is hence no
single point of failure for the entire DNS system. As today, caching
of generic domain name records by local DNS servers further increases
the robustness and performance for generic DNS lookups.

\smallskip \myparagraph{Security} The record for a generic domain~$D$
is signed by the owner of $D$ (i.e., the registrant~$r$) using an
asymmetric signature scheme and the private signing key of $r$. The
public verification key of $r$ is signed by $r$'s ISD and by the DNS
server for $\domain{.com}$. Client $C$ can hence base its trust on
the ISD of $r$ or (in case $C$ does not trust this ISD) on the DNS
server for $\domain{.com}$.

Another positive aspect of this design is the fact that a key
compromise of \domain{.com}'s DNS server does not directly affect the
security of an end-to-end connection: an attacker would additionally
need to compromise the DNS server of a regional ISD, which is used
for the second lookup, the regional lookup for the actual
address~$(\cI,A,E)$. Despite being unlikely, this attack only works
under the assumption that a client uses the verification key of
\domain{.com} (rather than the verification key of the resolved
regional CNAME domain).